\documentclass[aps,pre,twocolumn,superscriptaddress,superscriptaddress,longbibliography]{revtex4-2}
\usepackage{amsmath}
\usepackage{amsfonts}
\usepackage{amssymb}
\usepackage{lmodern,dsfont}
\usepackage{graphicx}
\usepackage[usenames,dvipsnames]{xcolor}
\usepackage{bm}
\usepackage[english=american]{csquotes}
\usepackage{xr}

\newcommand{\kb}{k_\text{B}}
\newcommand{\Av}[1]{\left\langle #1 \right\rangle}
\newcommand{\av}[1]{\langle #1 \rangle}
\newcommand{\n}{\nonumber}
\newcommand{\nn}{\nonumber \\}

\newcommand{\grad}{\bm{\nabla}}

\renewcommand{\eqref}[1]{Eq.~(\ref{#1})}
\newcommand{\eref}[1]{(\ref{#1})}

\begin{document}

\author{Andreas Dechant}
\affiliation{Department of Physics \#1, Graduate School of Science, Kyoto University, Kyoto 606-8502, Japan}
\author{Shin-ichi Sasa}
\affiliation{Department of Physics \#1, Graduate School of Science, Kyoto University, Kyoto 606-8502, Japan}
\title{Continuous time-reversal and equality in the thermodynamic uncertainty relation}
\date{\today}

\begin{abstract}
We introduce a continuous time-reversal operation which connects the time-forward and time-reversed trajectories in the steady state of an irreversible Markovian dynamics via a continuous family of stochastic dynamics.
This continuous time-reversal allows us to derive a tighter version of the thermodynamic uncertainty relation (TUR) involving observables evaluated relative to their local mean value.
Moreover, the family of dynamics realizing the continuous time-reversal contains an equilibrium dynamics halfway between the time-forward and time-reversed dynamics.
We show that this equilibrium dynamics, together with an appropriate choice of the observable, turns the inequality in the TUR into an equality.
We demonstrate our findings for the example of a particle diffusing in a tilted periodic potential.
\end{abstract}

\maketitle

The behavior of a system under time-reversal is one of its fundamental physical properties.
While most microscopic laws of physics are invariant under time-reversal, this is generally not true for macroscopic systems.
In addition to the energy-driven transitions between microscopic states, we also have to account for entropy, i.~e., the number of microscopic states that are compatible with a certain macroscopic state.
Thus, even if two macroscopic states are energetically equivalent, the likelihood of observing them may be vastly different, and the transitions from less likely to more likely macroscopic states lead to a breaking of time-reversal symmetry and an increase in entropy.

Irreversibility is made explicit in the framework of stochastic thermodynamics; there, the entropy production $\Delta S_\tau^\text{irr}$ during a time-interval $[0,\tau]$ is defined via the probabilities $\mathbb{P}_\tau(\Gamma)$ and $\mathbb{P}_\tau^\dagger(\Gamma)$ of observing a given trajectory $\Gamma$ of the system forward and time-reversed process, respectively \cite{Sek10,Sei12},
\begin{align}
\Delta S_\tau^\text{irr} = D_\text{KL}\big(\mathbb{P}_\tau \Vert \mathbb{P}_\tau^\dagger\big) = \int d\Gamma \ \mathbb{P}_\tau(\Gamma) \ln \bigg( \frac{\mathbb{P}_\tau(\Gamma)}{\mathbb{P}_\tau^\dagger(\Gamma)} \bigg) \label{entropy} .
\end{align}
$D_\text{KL}$ denotes the Kullback-Leibler (KL) divergence.
The entropy production is positive, except when the system is symmetric under time reversal, $\mathbb{P}_\tau(\Gamma) = \mathbb{P}_\tau^\dagger(\Gamma)$.
The definition \eqref{entropy} agrees with the thermodynamic definition of entropy for systems in contact with a heat bath, and also implies a stochastic entropy production along a single trajectory \cite{Sek10,Sei12},
\begin{align}
\Sigma_\tau(\Gamma) = \ln \bigg( \frac{\mathbb{P}_\tau(\Gamma)}{\mathbb{P}_\tau^\dagger(\Gamma)} \bigg) \label{entropy-stochastic},
\end{align}
such that its average is $\av{\Sigma_\tau} = \Delta S^\text{irr}_\tau$.

Intuitively, the entropy production $\Delta S_\tau^\text{irr}$ should also control to what degree physical observables can exhibit irreversibility.
This connection is made explicit in the thermodynamic uncertainty relation (TUR) \cite{Bar15,Gin16,Dec17,Pie17,Hor20}.
The TUR, which applies to steady states of irreversible Markovian dynamics, is an inequality between the average and fluctuations of an observable time-integrated current $J_\tau$ and the entropy production $\Delta S^\text{irr}_\tau$,
\begin{align}
\frac{\big(\av{J_\tau}\big)^2}{\text{Var}(J_\tau)} \leq \frac{1}{2} \Delta S_{\text{irr},\tau} \label{TUR}.
\end{align}
Here, $\av{J_\tau}$ denotes the average accumulated current up to time $\tau$ (see \eqref{current-average-0}) and $\text{Var}(J_\tau) = \av{J_\tau^2} - \av{J_\tau}^2$ is the variance.
The TUR is a tradeoff relation between precision and dissipation \cite{Shi16,Dec17,Pie18}: For a fixed average amount of physical quantity (particles, work, heat, ...) being transported, the product of fluctuations and dissipation cannot be less than the bound \eqref{TUR}; thus small fluctuations imply large dissipation. 

The TUR relates the statistics of an current, which is odd under time reversal, $\av{J_\tau}^\dagger = - \av{J_\tau}$, to the entropy production, which quantifies the asymmetry of the trajectories under time-reversal.
This suggests that this symmetry may be responsible for \eqref{TUR}.
A variant of the TUR, in which the right-hand side is proportional to the exponential of the entropy production, was derived from this symmetry in Ref.~\cite{Has19}.
However, this bound is generally less tight than \eqref{TUR} \cite{Fal20}.
In this Letter, we show that, indeed, the TUR is the consequence of the symmetry under a different type of time-reversal operation.

In general, time-reversal is a discrete operation, replacing the time-forward with the time-reversed process.
Our main result is that for the systems satisfying the TUR, there also exists a continuous time-reversal operation.
This operation describes a family of processes, parameterized by $\theta \in [-1,1]$, which connects the time-forward process at $\theta = 1$ to the time-reversed process at $\theta = -1$.
For any value of $\theta$, we have $\Sigma^\theta_\tau = \theta \Sigma_\tau$ and $\av{J_\tau}^\theta = \theta \av{J_\tau}$, so that the stochastic entropy production \eqref{entropy-stochastic} and symmetry of currents both extend in a natural way to the continuous case.
Further, every member of the family has the same steady state $p^{\theta}_\text{st} = p_\text{st}$.
Intuitively, the continuous time-reversal operation can be thought of as adiabatically changing the direction of the irreversible flows in the system: 
First, we reduce the magnitude of the flows while keeping the steady state fixed; at $\theta = 0$, the flows vanish and the system is in equilibrium. 
Then, we increase the magnitude of the flows in the opposite direction until at $\theta = -1$, all the flows have the same magnitude but opposite direction.

The continuous nature of this time-reversal operation allows us to derive tighter inequalities, as compared to a discrete operation.
Instead of an exponential bound \cite{Has19}, we obtain the linear inequality \eqref{TUR}.
As our second main result, we further obtain two variants of the TUR,
\begin{align}
\frac{\big(\av{J_\tau}\big)^2}{\text{Var}^0(J_\tau)} &\leq \frac{1}{2} \Delta S^\text{irr}_\tau, \label{TUR-eq} \\
\frac{\big(\av{J_\tau}\big)^2}{\text{Var}(\delta J_\tau)} &\leq \frac{1}{2} \Delta S^\text{irr}_\tau \label{RTUR}.
\end{align}
Compared to \eqref{TUR}, the difference is in the denominator on the left-hand side.
In \eqref{TUR-eq}, the fluctuations of the current are replaced by the fluctuations in the equilibrium process at $\theta = 0$. 
In \eqref{RTUR}, on the other hand, we consider the fluctuations $\delta J_\tau = J_\tau - \bar{J}_\tau$ relative local mean current $\bar{J}_\tau$, which is the current for a particle moving with the local mean velocity, see \eqref{current-splitting}.
We refer to \eqref{RTUR} as the relative TUR (RTUR).
In most cases of interest, we have $\text{Var}^0(J_\tau) < \text{Var}(J_\tau)$ and $\text{Var}(\delta J_\tau) < \text{Var}(J_\tau)$ and both \eqref{TUR-eq} and \eqref{RTUR} are tighter inequalities than the TUR \eref{TUR}.
In particular, both \eqref{TUR-eq} and \eqref{RTUR} reduce to an equality when we choose the stochastic entropy production as the observable, $J_\tau = \Sigma_\tau$.
This is in contrast to the TUR, which reduces to the Fano-factor inequality derived in Ref.~\cite{Pig17}.

\textit{Continuous time-reversal.} 
For simplicity, we focus on the case of an overdamped Langevin dynamics in the following.
We consider a system of $N$ degrees of freedom $\bm{x}(t) = (x_1(t),\ldots,x_N(t))$ whose motion is described by the overdamped Langevin equation during the time-interval $t \in [0,\tau]$ \cite{Ris86}
\begin{align}
\dot{\bm{x}}(t) = \bm{a}(\bm{x}(t)) + \bm{G} \cdot \bm{\xi}(t) \label{langevin} .
\end{align}
Here, $\bm{a}(\bm{x})$ is the drift vector, and we assume that the matrix $\bm{G}$ has full rank such that the diffusion matrix $\bm{B} = \bm{G} \bm{G}^\text{T}/2$, where the superscript T denotes transposition, is positive definite.
$\bm{\xi}(t)$ is a vector of uncorrelated Gaussian white noises.
The extension to a coordinate dependent matrix $\bm{G}(\bm{x})$ is provided in the Supplemental Material (SM) \cite{supmat}.
The paradigmatic example is a system of $N$ particles with systematic forces $\bm{f}(\bm{x})$, which diffuse in an environment described by a mobility $\mu$ and a temperature $T$.
In this case, we have $\bm{a}(\bm{x}) = \mu \bm{f}(\bm{x})$ and $\bm{B} = \mu \kb T \bm{1}$.
We assume that $\bm{a}(\bm{x})$ and $\bm{B}$ give rise to a time-independent state in the long-time limit, i.~e.~that the solution of the associated Fokker-Planck equation for the probability density $p(\bm{x},t)$ \cite{Ris86},
\begin{align}
\partial_t p(\bm{x},t) &= - \grad \cdot \big(\bm{\nu}(\bm{x},t) p(\bm{x},t) \big)   \label{fokkerplanck} \qquad \text{with} \\
\bm{\nu}(\bm{x},t) &= \bm{a}(\bm{x}) - \bm{B} \grad \ln p(\bm{x},t) \n ,
\end{align}
tends, as $t \rightarrow \infty$, towards a steady state solution $p_\text{st}(\bm{x})$ with local mean velocity $\bm{\nu}_\text{st}(\bm{x})$.
Physically, the local mean velocity $\bm{\nu}_\text{st}(\bm{x})$ characterizes the irreversible local flows in the system \cite{Hat01,Spe06}.
Since, generally, the system described by \eqref{langevin} is out of equilibrium, these flows do not vanish even in the steady state.
We use the local mean velocity to write the drift vector as 
\begin{align}
\bm{a}(\bm{x}) = \bm{\nu}_\text{st}(\bm{x}) + \bm{B} \grad \ln p_\text{st}(\bm{x}) \label{meanvel} .
\end{align} 
\eqref{meanvel} may be viewed as a decomposition of the drift vector into an irreversible part $\bm{\nu}_\text{st}(\bm{x})$ and a reversible part \cite{Hat01,Spe06,Sei10,Spi12,Sas14}.
We introduce a modified drift vector,
\begin{align}
\bm{a}^\theta(\bm{x}) = \theta \bm{\nu}_\text{st}(\bm{x}) + \bm{B} \grad \ln p_\text{st}(\bm{x}),
\end{align}
with a parameter $\theta \in [-1,1]$, and consider the corresponding Langevin dyamics
\begin{align}
\dot{\bm{x}}(t) = \bm{a}^\theta(\bm{x}(t)) + \bm{G} \cdot \bm{\xi}(t) . \label{langevin-mod}
\end{align}
Compared to \eqref{meanvel}, we have rescaled the irreversible part of the drift vector, while leaving the reversible part unchanged.
It is straightforward to verify that the steady-state solution for \eqref{langevin-mod} is given by $p^\theta_\text{st}(\bm{x}) = p_\text{st}(\bm{x})$ and $\bm{\nu}_\text{st}^\theta(\bm{x}) = \theta \bm{\nu}_\text{st}(\bm{x})$, i.~e.~we obtain the same steady-state density as \eqref{langevin} and a local mean velocity scaled by a factor $\theta$.
The family of dynamics \eqref{langevin-mod} was previously studied in Ref.~\cite{Che06}, where it was shown to lead to generalized fluctuation theorems.
Here and in the following, we use a superscript $\theta$ to refer to quantities evaluated in the dynamics with drift vector \eqref{langevin-mod}; quantities without a superscript refer to \eqref{langevin}.
For each value of $\theta$, \eqref{langevin-mod} generates a path probability density $\mathbb{P}_\tau^\theta[\hat{\bm{x}}]$, which measures the probability of observing a specific trajectory $\hat{\bm{x}} = (\bm{x}(t))_{t \in [0,\tau]}$.
For each trajectory, we can also consider its time-reversed version $\hat{\bm{x}}^\dagger = (\bm{x}(\tau-t))_{t \in [0,\tau]}$.
For the dynamics \eqref{langevin-mod} in the steady state, this time-reversed trajectory defines the path probability of the reverse process, $\mathbb{P}_\tau^{\theta,\dagger}[\hat{\bm{x}}] = \mathbb{P}_\tau^\theta[\hat{\bm{x}}^\dagger]$.
A technical but straightforward calculation (see Eq.~(S50) of the SM \cite{supmat}) shows that the time-reversed path probability satisfies
\begin{align}
D_\text{KL} \big( \mathbb{P}_\tau^{-\theta}[\hat{\bm{x}}] \hspace{.25mm} \big\Vert \hspace{.25mm} \mathbb{P}_\tau^{\theta,\dagger}[\hat{\bm{x}}] \big) = 0 \label{KL-time-reverse}.
\end{align}
If the KL divergence between two probability densities vanishes, then the two probability densities are equivalent: any average evaluated with respect to either of them yields the same result.
From this, we can conclude that the dynamics \eqref{langevin-mod} at $-\theta$ is equivalent to the time-reversed dynamics at $\theta$.
In particular, \eqref{langevin-mod} for $\theta = -1$ yields the time-reversed dynamics of \eqref{langevin}, see Ref.~\cite{Sas14}.
Thus, for a general non-equilibrium dynamics, \eqref{langevin-mod} provides a continuous interpolation between the original, time-forward dynamics for $\theta = 1$ and the time-reversed dynamics for $\theta = -1$.
For $\theta = 0$, the irreversible part of the drift vanishes and \eqref{langevin-mod} describes an equilibrium system.
However, this does not necessarily correspond to the intuitive, \enquote{physical} equilibrium.
The reason is that, when driving a system out of equilibrium by applying a non-conservative force, the state density $p_\text{st}(\bm{x})$ is generally different from the equilibrium state $p_\text{eq}(\bm{x})$ in the absence of the driving.
By contrast, for \eqref{langevin-mod} with $\theta = 0$, the steady state $p_\text{st}(\bm{x})$ is the equilibrium state.

\textit{Thermodynamic uncertainty relation.}
While \eqref{KL-time-reverse} provides a relation between the time-reversed dynamics at $-\theta$ and the time-forward dynamics at $\theta$, we also obtain a relation between the time-forward dynamics at two different values of $\theta$ (see Eq.~(S32) of the SM \cite{supmat}),
\begin{align}
D_\text{KL} \big( \mathbb{P}_\tau^{\theta}[\hat{\bm{x}}] \hspace{.25mm} \big\Vert \hspace{.25mm} \mathbb{P}_\tau^{\theta'}[\hat{\bm{x}}] \big) = \frac{1}{4}\big(\theta' - \theta \big)^2 \Delta S_{\text{irr},\tau} \label{KL-general}.
\end{align}
For $\theta = 1$ and $\theta' = -1$, this is precisely \eqref{entropy}.
Surprisingly, the entropy production not only characterizes the difference between the forward and reverse dynamics, but also between any two members of the family of dynamics \eqref{langevin-mod}.
Next, we establish the connection between \eqref{langevin-mod} and time-integrated currents.
The latter are defined as
\begin{align}
J_\tau = \int_0^\tau dt \ \bm{w}(\bm{x}(t)) \circ \dot{\bm{x}}(t) \label{current} ,
\end{align}
where $\bm{w}(\bm{x})$ is a weighting function and $\circ$ is the Stratonovich product.
If $\bm{w}(\bm{x}) = \bm{e}$ is a constant vector of unit length, then $J_\tau$ is the displacement along the direction $\bm{e}$.
Another physically relevant choice is $\bm{w}(\bm{x}) = \bm{f}(\bm{x})$, in which case $J_\tau$ is the heat dissipated into the surrounding environment.
The steady-state average of \eqref{current} is given by
\begin{align}
\av{J_\tau} = \tau \int d \bm{x} \ \bm{w}(\bm{x}(t)) \cdot \bm{\nu}_\text{st}(\bm{x}) p_\text{st}(\bm{x}) \label{current-average-0}.
\end{align}
Since this is proportional to the local mean velocity, the average current in the dynamics with \eqref{langevin-mod} exhibits the same scaling,
\begin{align}
\av{J_\tau}^\theta = \theta \av{J_\tau}, \label{current-average}
\end{align}
and we have $\av{J_\tau}^0 = 0$ and $\av{J_\tau}^{-1} = -\av{J_\tau}$; the average current vanishes in equilibrium and time reversal changes its sign.
Now, we return to \eqref{KL-general} and focus on the case $\theta' = \theta + d\theta$ with $d\theta \ll 1$.
Using the fluctuation-response inequality for linear response derived in Ref.~\cite{Dec20}, we have 
\begin{align}
\frac{\big(\av{J_\tau}^{\theta + d\theta} - \av{J_\tau}^\theta \big)^2}{2 \text{Var}^\theta(J_\tau)} \leq D_\text{KL} \big( \mathbb{P}^{\theta}_\tau[\hat{\bm{x}}] \big\Vert \mathbb{P}^{\theta + d\theta}_\tau[\hat{\bm{x}}] \big) \label{FRI} .
\end{align}
Using \eqref{KL-general} and \eqref{current-average}, this yields
\begin{align}
\frac{\big(\av{J_\tau}\big)^2}{\text{Var}^\theta(J_\tau)} \leq \frac{1}{2} \Delta S^\text{irr}_{\tau} \label{TUR-theta} .
\end{align}
Since this is valid for any value of $\theta \in [-1,1]$, we may also maximize the left-hand side over $\theta$, which yields 
\begin{align}
\frac{\big(\av{J_\tau}\big)^2}{\inf_{\theta}\big(\text{Var}^\theta(J_\tau)\big)} \leq \frac{1}{2} S^\text{irr}_{\tau} \label{TUR-minvar} .
\end{align}
This bound is tighter than \eqref{TUR}; further, any value of $\theta$ yields a valid bound.
In particular, we may choose $\theta = 1$ and obtain \eqref{TUR} or $\theta = 0$ and obtain \eqref{TUR-eq}.
We remark that \eqref{TUR-theta} is conceptually different from previous formulations of the TUR, since it relates observables evaluated in different dynamics.

The variance is the second cumulant of the current.
However, if the distribution of the current is not Gaussian, the current also possesses non-vanishing higher-order cumulants.
These can be calculated from the cumulant generating function
\begin{align}
K_{J_\tau}^\theta(h) = \ln \int d\hat{\bm{x}} \ e^{h J_\tau[\hat{\bm{x}}]} \mathbb{P}_\tau^\theta[\hat{\bm{x}}],
\end{align}
in terms of which the $n$-th cumulant $\kappa_{J_\tau}^{(n),\theta}$ is defined as ${\partial_h}^n K_{J_\tau}^\theta(h) \vert_{h = 0}$.
Since the currents are odd under time-reversal, this satisfies
\begin{align}
K_{J_\tau}^{-\theta}(h) = K_{J_\tau}^{\theta}(-h),
\end{align}
that is, even cumulants are invariant under the change $\theta \rightarrow -\theta$, while odd cumulants change sign.
This implies
\begin{align}
\kappa_{J_\tau}^{(n),0} = 0 \quad \text{for} \; n \; \text{odd},
\end{align}
all odd cumulants vanish in the equilibrium state at $\theta = 0$.
As demonstrated in Eq.~(S84) of the SM \cite{supmat}, we may also use the higher-order cumulants to obtain a generalization of \eqref{TUR-minvar},
\begin{align}
\Delta S^\text{irr}_{\tau} \geq \sup_{h,\theta} \bigg( \frac{h^2 \big(\av{J_\tau} \big)^2 }{K_{J_\tau}^{\theta}(h) - h \theta \av{J_\tau}} \bigg) .
\end{align}
This reduces to \eqref{TUR-minvar} in the limit $h \rightarrow 0$, but yields a tighter bound if the higher-order cumulants of the current are known.
In particular, for $\theta = 1$, we obtain a higher-order TUR,
\begin{align}
\Delta S^\text{irr}_{\tau} \geq \sup_{h} \bigg( \frac{h^2 \big(\av{J_\tau} \big)^2 }{K_{J_\tau}(h) - h \av{J_\tau}} \bigg) .
\end{align}

\textit{Current fluctuations.}
We define the local mean value $\bar{J}_\tau$ of the current \eqref{current} by replacing the velocity with its local mean value,
\begin{align}
\bar{J}_\tau &= \int_0^\tau dt \ \bm{w}(\bm{x}(t)) \cdot \bm{\nu}_\text{st}(\bm{x}(t))  \label{current-splitting} ,
\end{align}
and the current relative to the local mean value $\delta J_\tau = J_\tau - \bar{J}_\tau$.
From the definition, it is clear that $\av{\bar{J}_\tau} = \av{J_\tau}$ and $\av{\delta J_\tau} = 0$, i.~e.~only $\bar{J}_\tau$ contributes to the average current.
Evaluating the average of $\delta J_\tau$ in the dynamics \eqref{langevin-mod}, we obtain
\begin{align}
\av{\delta J_\tau}^\theta = (\theta - 1) \av{J_\tau} \; \Rightarrow \; \av{\delta J_\tau}^{\theta + d\theta} - \av{\delta J_\tau}^\theta = d\theta \av{J_\tau} .
\end{align}
Using this in \eqref{FRI}, we obtain the inequality
\begin{align}
\frac{\big( \av{J_\tau} \big)^2}{\text{Var}^\theta(\delta J_\tau)} \leq \frac{1}{2} S^\text{irr}_{\tau}.
\end{align}
For $\theta = 1$, we find the RTUR \eref{RTUR} involving the current relative to its local mean value.
Generally, $\text{Var}(\delta J_\tau)$ may be larger or smaller than $\text{Var}(J_\tau)$, and thus, either the TUR or the RTUR may be tighter.
However, the relation $\text{Var}(J_\tau) \geq \text{Var}(\delta J_\tau)$ often holds in practice, where \eqref{RTUR} thus provides a tighter bound than \eqref{TUR}.
We provide an example for this behavior below.

\textit{Entropy fluctuations.}
An important case of a time-integrated current \eqref{current} is $\bm{w}(\bm{x}) = \bm{B}^{-1} \bm{\nu}_\text{st}(\bm{x})$, for which $J_\tau = \Sigma_\tau$ is equal to the stochastic entropy production \eqref{entropy-stochastic},
\begin{align}
\Sigma_\tau[\hat{\bm{x}}] &= \ln \frac{\mathbb{P}_\tau[\hat{\bm{x}}]}{\mathbb{P}_\tau^\dagger[\hat{\bm{x}}]} \label{entropy-stochastic-2} . 
\end{align}
The equivalence between \eqref{current} with $\bm{w}(\bm{x})$ as above and \eqref{entropy-stochastic-2} is established in section I.B. of the SM \cite{supmat}.
Written in this way $\Sigma_\tau$ explicitly depends on the path statistics of the entire ensemble. 
For the dynamics \eqref{langevin-mod} we may similarly write
\begin{align}
\Sigma_\tau^\theta[\hat{\bm{x}}] = \ln \frac{\mathbb{P}^\theta_\tau[\hat{\bm{x}}]}{\mathbb{P}_\tau^{\theta,\dagger}[\hat{\bm{x}}]} .
\end{align}
Using the definition of $\Sigma_\tau^\theta$ in terms of $\bm{w}(\bm{x}) = \bm{B}^{-1} \bm{\nu}^\theta_\text{st}(\bm{x})$ together with the scaling of the local mean velocity $\bm{\nu}^\theta_\text{st}(\bm{x}) = \theta \bm{\nu}_\text{st}(\bm{x})$, we immediately find
\begin{align}
\Sigma_\tau^\theta[\hat{\bm{x}}] = \theta \Sigma_\tau[\hat{\bm{x}}] \label{entropy-scaling} .
\end{align}
This means that the parameter $\theta$ determines the relative likelihood of observing a trajectory as a forward or reverse trajectory in the dynamics with \eqref{langevin-mod}, with both possibilities being equally likely at $\theta = 0$.
Evaluating the variance of $\Sigma_\tau$ (see Ref.~\cite{Pig17} and Eq.~(S72) of the SM \cite{supmat}), we find
\begin{align}
\text{Var}(\Sigma_\tau) = \text{Var}(\bar{\Sigma}_\tau) + \text{Var}(\delta \Sigma_\tau) \label{entropy-splitting}.
\end{align}
Formally, \eqref{entropy-splitting} is equivalent to the introduction of an entropic time in Ref.~\cite{Pig17}:
The quantity $\bar{\Sigma}_\tau$ can be interpreted as a dimensionless, stochastic time coordinate.
Then, $\delta \Sigma_\tau$ is equal to the entropy production measured in units of the stochastic time.
As was shown in Ref.~\cite{Pig17}, this implies that the distribution of $\delta \Sigma_\tau$ is Gaussian.
Further, we have the identities
\begin{align}
\text{Var}(\delta \Sigma_\tau) = 2 \Delta S^\text{irr}_{\tau} = \text{Var}^0(\Sigma_\tau) \label{entropy-fluctuations} .
\end{align}
Comparing the first identity to the RTUR \eref{RTUR}, we see that the latter turns into an equality.
The second identity turns \eqref{TUR-eq} into an equality.
Using this, we may write the variational expression
\begin{align}
\sup_{J,\theta} \Bigg( \frac{\big(\av{J_\tau} \big)^2}{\text{Var}^\theta(J_\tau)} \Bigg) = \sup_{J} \Bigg( \frac{\big(\av{J_\tau} \big)^2}{\text{Var}(\delta J_\tau)} \Bigg) = \frac{1}{2} \Delta S^\text{irr}_{\tau} \label{TUR-variational} ,
\end{align}
which characterizes the equality condition for the TUR \eref{TUR} and the RTUR \eref{RTUR}.
Close to equilibrium, we have $\text{Var}(\Sigma_\tau) \simeq \text{Var}(\delta \Sigma_\tau)$ and the TUR turns into an equality by choosing the stochastic entropy production as an observable \cite{Pie16,Mac18}.
Indeed, the relation $\text{Var}(\Sigma_\tau) \simeq 2 S^\text{irr}_{\tau}$ follows from the fluctuation-dissipation theorem \cite{Web56,Kub66}.
Far from equilibrium, this breaks down, and there is generally no observable that turns the TUR into an equality; to realize the equality, we have to replace the current fluctuations with their equilibrium value at $\theta = 0$.
This suggests that the presence of excess fluctuations out of equilibrium prohibits equality in the TUR.
On the other hand, equality in the RTUR \eref{RTUR} may always be realized by choosing the stochastic entropy production as an observable.
Just like the velocity relative to the local mean velocity recovers the equilibrium fluctuation-dissipation theorem \cite{Spe06}, the current relative to the local mean value recovers the equilibrium equality condition for the TUR.

\begin{figure}
\includegraphics[width=.47\textwidth]{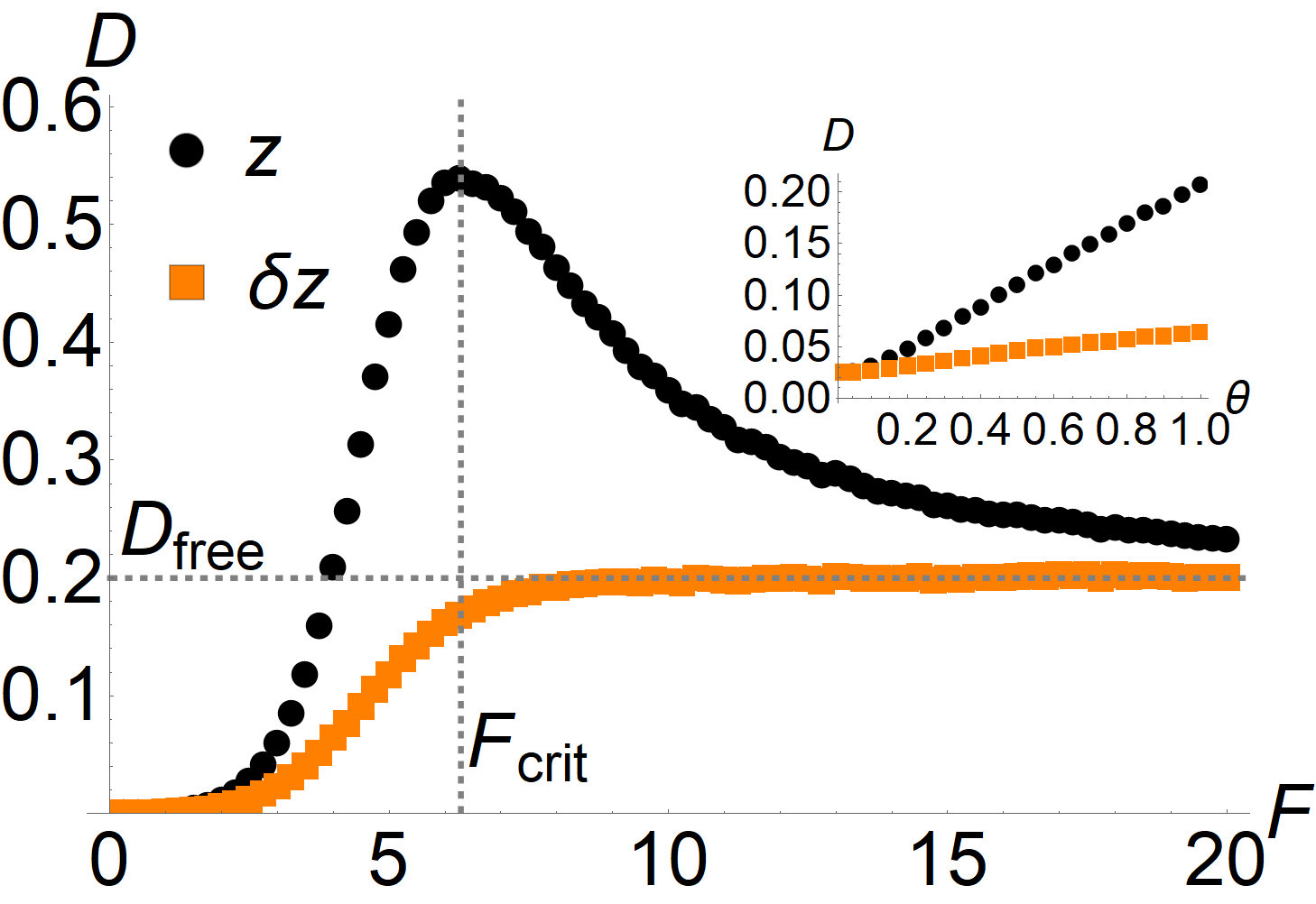}\\
\includegraphics[width=.47\textwidth]{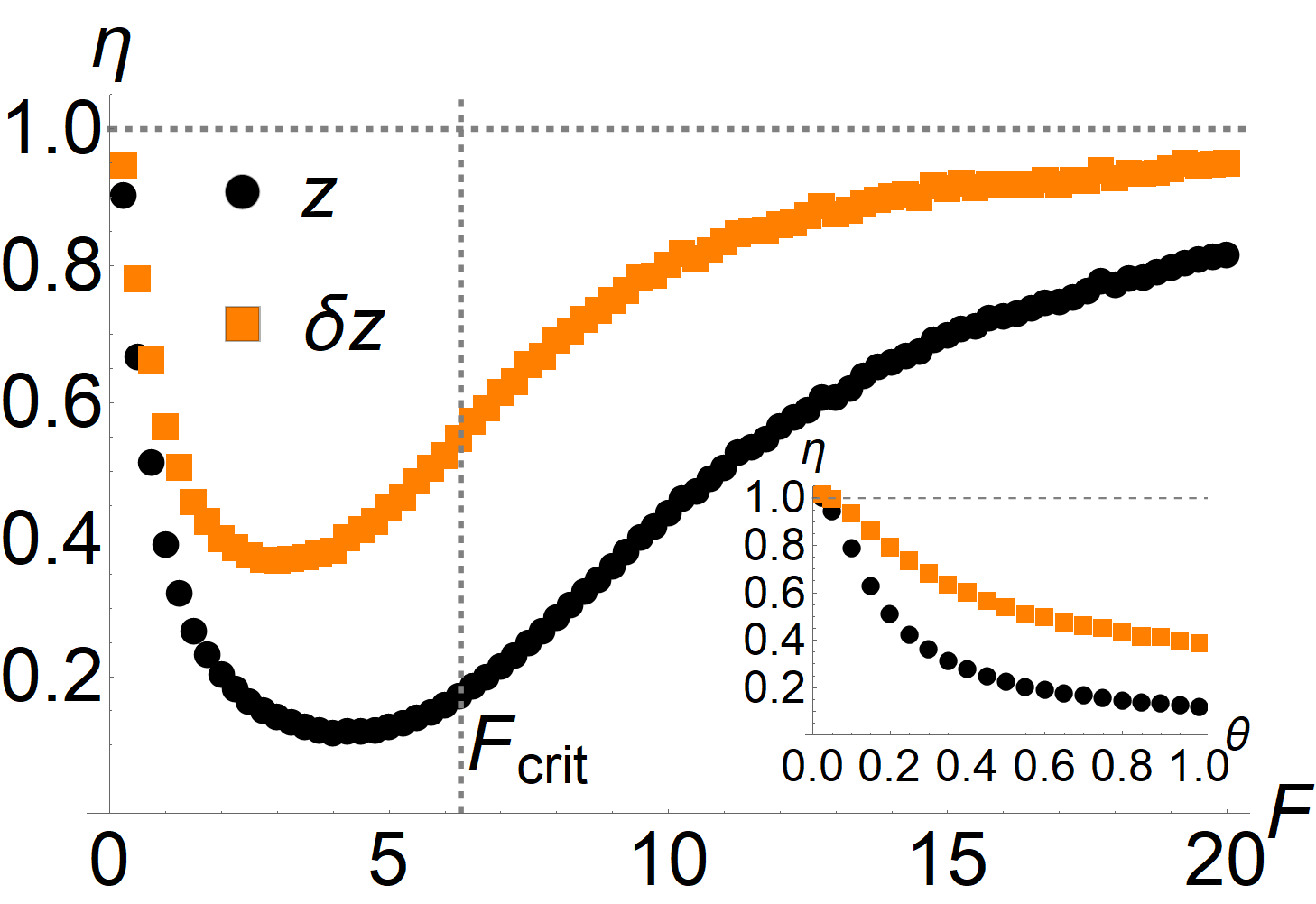}
\caption{The diffusion coefficient (top) and the transport efficiency (bottom) as a function of the bias force (main panel) and the parameter $\theta$ for $F = 4$ (inset). Black dots show the respective quantity for the displacement, while the orange squares correspond to the fluctuations of the displacement around its local mean value. The data was obtained using Langevin simulations in a sine-potential $U(x) = U_0 \sin( 2 \pi x/L)$ with $U_0 = 1$, $L = 1$. The temperature and mobility were set to $T = 0.2$ and $\mu = 1$. \label{fig-0}}
\end{figure}

\textit{Demonstration: Tilted periodic potential.}
We illustrate our results using a paradigmatic example of a non-equilibrium steady state.
We consider a Brownian particle in one dimension with mobility $\mu$ and at temperature $T$, which moves in a periodic potential $U(x+L) = U(x)$.
This situation is described by the Langevin equation
\begin{align}
\dot{x}(t) = \mu( -U'(x(t)) + F) + \sqrt{2 \mu \kb T} \xi(t) .
\end{align}
The system is driven out of equilibrium by the constant bias force $F$. 
The (periodic) steady state probability density and local mean velocity for this system may be computed explicitly, see \cite{Rei01} and Section IV of the SM \cite{supmat}.
Since the steady state probability density differs from the Boltzmann-Gibbs density $p_\text{eq}(x) \propto e^{-U(x)/(\kb T)}$, the equilibrium state for $\theta = 0$ does not coincide with the physical equilibrium at $F = 0$.
In the following, we focus on the displacement $z_\tau$ of the particle with $w(x) = 1$ in \eqref{current}.
In the long-time limit, displacement behaves diffusively $\text{Var}(z_\tau) \simeq 2 D_z \tau$; an explicit expression for the diffusion coefficient $D_z$ was derived in Ref.~\cite{Rei01}.
One remarkable feature appears for low temperatures and a bias force close to the critical value $F_\text{crit}$, at which the minima of the tilted potential disappear. 
Under these conditions, the diffusion coefficient can be orders of magnitude larger than the free diffusion coefficient in the absence of the periodic potential $D_{z,\text{free}} = \mu \kb T$ \cite{Rei01}.
As a function of the bias, the diffusion coefficient is small for small bias, reaches a maximum near critical tilt and then decreases towards the free value, see Fig.~\ref{fig-0}.
However, this enhancement of diffusion is absent in the displacement relative to the local mean value: The corresponding diffusion coefficient $D_{\delta z}$ increases monotonously towards the free value and is always smaller than $D_z$, see Fig.~\ref{fig-0}.
As a consequence, we have
\begin{align}
\eta_z = \frac{2 \big(\av{z_\tau}\big)^2}{\text{Var}(z_\tau) S^\text{irr}_{\tau}} \leq \eta_{\delta z} = \frac{2 \big(\av{z_\tau}\big)^2}{\text{Var}(\delta z_\tau) S^\text{irr}_{\tau}} \leq 1,
\end{align}
i.~e.~the RTUR \eref{RTUR} is tighter than the TUR \eref{TUR}.
For small bias (near equilibrium), $\eta_z$ approaches unity.
For large bias, the potential becomes negligible and the system behaves like biased diffusion, where $\eta_z$ likewise approaches unity.
For intermediate bias, on the other hand, $\eta_z$ is significantly smaller than unity.
In this regime, the bound involving $\delta z_\tau$ is considerably tighter, indicating that the decrease in $\eta_z$ is partly due to the enhancement of the diffusion coefficient.
Note that, in general, the definition of the local mean current \eqref{current-splitting} involves the local mean velocity and may thus be difficult to compute in cases where the latter is not explicitly known.
However, for one-dimensional systems, we have the relation
\begin{align}
\nu_\text{st}(x) = \frac{\av{\dot{z}}}{L p_\text{st}(x)},
\end{align}
and thus $\nu_\text{st}(x)$ and $\delta z$ can be evaluated by measuring the steady-state probability density.
Finally, we remark that, while both $F = 0$ and $\theta = 0$ (for finite $F$) correspond to an equilibrium dynamics, the non-monotonic behavior in $D_z$ and $\eta_z$ only appears as a function of $F$.
By contrast, $D_z$ ($\eta_z$) increases (decreases) monotonically when changing $\theta$ from $0$ to $1$, see the insets of Fig.~\ref{fig-0}.

\textit{Discussion.}
The dynamics \eqref{langevin-mod} provide a natural way to interpolate between the time-forward and the time-reversed dynamics, replacing a discrete operation with a continuous one.
A continuous operation can be represented by a series of infinitesimal steps, which can then be analyzed individually, reconstructing the entire operation from the individual steps.
In the present context, this allows us to apply the linear-response fluctuation-response inequality \eqref{FRI}, providing a tighter inequality than can be obtained by directly comparing the time-forward and time-reversed process (see also Section III of the SM \cite{supmat}).

In many applications, non-equilibrium states are obtained by driving an equilibrium system, for example by applying non-conservative force.
In this case, the non-equilibrium system has a natural equilibrium counterpart.
However, for a given non-equilibrium state, this equilibrium is not unique; the same non-equilibrium state may be obtained by driving two different systems in different ways.
Thus, knowledge about the equilibrium system may not necessarily tell us anything about the non-equilibrium state.
By contrast, the continuous time-reversal operation continuously connects a non-equilibrium steady state to a unique equilibrium system with the same steady state.
As demonstrated in the insets of Fig.~\ref{fig-0}, the physical properties of the system change in a much more controlled fashion between this unique equilibrium and the non-equilibrium state, when compared with the physical equilibrium state. 
If this type of behavior can be shown to be generic, this may provide a new approach of characterizing non-equilibrium states in terms of equilibrium states and their well-understood properties.

A practical application of the TUR \eqref{TUR} is to estimate the entropy production and thus dissipation by measuring a current in the system \cite{Li19,Man20,Ots20,Vu20}.
Since the dissipation is often not directly accessible in experiments, relating it to measurable quantities is crucial.
Then, an obvious question is how good the lower estimate \eqref{TUR} on the entropy production can be.
The generally tighter bounds \eqref{TUR-eq} and \eqref{RTUR} restrict the quality of this estimate in terms of the fluctuations of the current.
If we have $\text{Var}(J_\tau) \geq \text{Var}^0(J_\tau), \text{Var}(\delta J_\tau)$, then this immediately implies that the estimate from the TUR will be too small by at least this amount.

While in this work, we focused on overdamped Langevin dynamics, the notion of continuous time-reversal and the results of this Letter also apply to Markov jump dynamics, as we will discuss in an upcoming publication \cite{Dec20b}.
Since the TUR follows explicitly as a consequence of the continuity, we speculate that finding a continuous time-reversal symmetry may serve as a way to extend the TUR to other classes of dynamics.
Whether such an operation exists depends on the dynamics; for example, it is known that the TUR can violated in the presence of magnetic fields which transform in a discrete manner under time-reversal \cite{Chu19}.
Similarly, it would be interesting to explore whether recent extensions of the TUR to non-steady initial states \cite{Liu20}, time-periodic \cite{Bar18,Koy18,Koy19} or arbitrary time-dependent driving \cite{Koy20} can be connected to the existence of a generalized continuous time-reversal operation.

\begin{acknowledgments}
\textbf{Acknowledgments.} This work was supported by KAKENHI (Nos. 17H01148, 19H05795 and 20K20425).
\end{acknowledgments}


%

\clearpage

\onecolumngrid

\begin{center}
\huge{Supplemental material}
\end{center}

\section{Path probability density and continuous time-reversal}

\subsection{Forward and time-reverse path probability for Langevin dynamics}
In the main text, we focused on a Langevin dynamics with position-independent diffusion matrix.
For the sake of generality, we include the possibility of a state-dependent diffusion matrix in the following.
The corresponding Langevin equation reads
\begin{align}
\dot{\bm{x}}(t) = \bm{a}(\bm{x}(t)) + \bm{G}(\bm{x}(t)) \cdot \bm{\xi}(t) \bm{(},
\end{align}
where $\bm{\xi}(t)$ is a vector of mutually independent standard Gaussian white noises and $\cdot$ denotes the Ito-product. 
As in the coordinate-independent case $\bm{G}(\bm{x})$ is assumed to have full rank for any $\bm{x}$.
The diffusion matrix is then the symmetric, positive definite matrix $\bm{B}(\bm{x}) = \bm{G}(\bm{x}) \bm{G}(\bm{x})^\text{T}$.
In \eqref{main-langevin-mod} of the main text, we introduced the parameterized Langevin dynamics
\begin{align}
\dot{\bm{x}}(t) = \underbrace{\theta \bm{\nu}_\text{st}(\bm{x}(t)) + \bm{\phi}_\text{st}(\bm{x}(t))}_{\equiv \bm{a}^\theta(\bm{x}(t))} + \bm{G}(\bm{x}(t)) \cdot \bm{\xi}(t) \label{langevin-mod},
\end{align}
with parameter $\theta \in [-1,1]$.
Here, $\bm{\nu}_\text{st}(\bm{x})$ is the steady state local mean velocity and we defined the reversible part of the drift vector
\begin{align}
\bm{\phi}_\text{st}(\bm{x}) = \big(\grad^\text{T} \bm{B}(\bm{x}) \big)^\text{T} - \bm{B}(\bm{x}) \grad \ln p_\text{st}(\bm{x}) .
\end{align}
The path probability density for this process can be constructed from the short-time solution of the associated Fokker-Planck equation for the transition probability density \cite{Ris86}
\begin{align}
\partial_t p^\theta(\bm{x},t \vert \bm{y},s) &= - \grad^\text{T} \Big( \big(\bm{a}^\theta(\bm{x}) - \grad \bm{B}(\bm{x}) \big) p(\bm{x},t \vert \bm{y},s) \Big),
\end{align}
where we defined the vector operator $\grad \bm{B}(\bm{x})$ with components $(\grad \bm{B}(\bm{x}) f(\bm{x}))_i = \partial_{x_j} ( B_{i j}(\bm{x}) f(\bm{x}))$ .
In the limit of small time-differences $dt \rightarrow 0$, the solution is given by the Gaussian propagator
\begin{align}
p^\theta_0(\bm{x},t+dt \vert \bm{y},t) \simeq \frac{1}{\sqrt{(4 \pi dt)^N \det(\bm{B}(y))}} \exp \bigg[ - \frac{1}{4 dt} \big( \bm{x} - \bm{y} - \bm{a}^\theta(\bm{y}) dt \big)^\text{T} \bm{B}^{-1}(\bm{y}) \big( \bm{x} - \bm{y} - \bm{a}^\theta(\bm{y}) dt \big) \bigg] \label{propagator} ,
\end{align}
where $N$ is the dimension of $\bm{x}$.
Since \eqref{langevin-mod} has no explicit time-dependence, we may drop the reference to $t$ and define
\begin{align}
p^\theta_{dt,0}(\bm{x} \vert \bm{y}) \equiv p^\theta_0(\bm{x},t+dt \vert \bm{y},t)
\end{align}
Note that here, we choose to write the propagator in initial-point discretization, i.~e.~we evaluate the drift vector and diffusion matrix at the coordinate $\bm{y}$ associated with the initial time $t$.
We may also write the propagator by evaluating these terms at $\bm{z} = (1-\alpha) \bm{y} + \alpha \bm{x}$ with $\alpha \in [0,1]$, which leads to the propagator \cite{Spi12}
\begin{align}
p^\theta_{dt,\alpha}(\bm{x} \vert \bm{y}) &\simeq \frac{1}{\sqrt{(4 \pi dt)^N \det(\bm{B}(z))}} \label{propagator-alpha} \\
&\qquad \times \exp \bigg[ - \frac{1}{4 dt} \Big[ \bm{x} - \bm{y} - \bm{a}^\theta(\bm{z}) dt - 2\alpha \grad \bm{B}(\bm{z}) dt \Big]^\text{T} \bm{B}^{-1}(\bm{z}) \Big[ \bm{x} - \bm{y} - \bm{a}^\theta(\bm{z}) dt - 2\alpha \grad \bm{B}(\bm{z}) dt \Big] \nn
& \hspace{2cm} - \alpha \grad^\text{T} \bm{a}^\theta(\bm{y}) dt + \alpha^2 \mathcal{H}_B(\bm{z}) dt \bigg] \n ,
\end{align}
where we defined
\begin{align}
\mathcal{H}_B(\bm{z}) = \sum_{i,j} \partial_{x_i} \partial_{x_j} \bm{B}_{i j}(\bm{x}) \Big\vert_{\bm{x} = \bm{z}} .
\end{align}
Here, we adopted the convention that derivatives enclosed in brackets only act on terms inside, e.~g.~$[\grad f(\bm{x})] g(\bm{x}) = g(\bm{x}) \grad f(\bm{x})$ while $\grad f(\bm{x}) g(\bm{x}) = f(\bm{x}) \grad g(\bm{x}) +  g(\bm{x}) \grad f(\bm{x})$.
The additional terms in the exponent reflect the requirement that $p^\theta_\alpha(\bm{x},t+dt \vert \bm{y},t)$ should be a normalized probability density with respect to $\bm{x}$ to leading order in $dt$, i.~e.~$\int d\bm{x} \ p^\theta_\alpha(\bm{x},t+dt \vert \bm{y},t) = 1 + O(dt^2)$ (note that for $\alpha \neq 0$, expression also depends on $\bm{x}$ implicitly via $\bm{z}$).
The propagator for any choice of $\alpha$ is equivalent to leading order in $dt$, however, the choice of $\alpha$ becomes relevant for the reverse propagator (see below).
From \eqref{propagator}, we construct the probability density of a discretized trajectory $\hat{\bm{x}}_K = (\bm{x}_K, \bm{x}_{K-1},\ldots,\bm{x}_1,\bm{x}_0)$ with $K = \tau/dt$ steps and $\tau$ being the length of the observation time-interval.
This is given by
\begin{align}
\mathbb{P}^\theta_K[\hat{\bm{x}}] = \prod_{k = 1}^K p_{dt,0}^\theta(\bm{x}_k \vert \bm{x}_{k-1}) p^\theta(\bm{x}_0,0),
\end{align}
where $p(\bm{x}_0,0)$ is the (given) initial probability density.
From this, the continuous-time path probability density is defined as the corresponding limit
\begin{align}
\mathbb{P}^\theta_\tau[\hat{\bm{x}}] = \lim_{dt \rightarrow 0} \mathbb{P}^\theta_K[\hat{\bm{x}}]
\end{align}
while keeping $\tau = K dt$ fixed.
As discussed in Ref.~\cite{Spi12}, when defining the stochastic entropy production as
\begin{align}
\Sigma^\theta_\tau = \ln \frac{\mathbb{P}^\theta_\tau[\hat{\bm{x}}]}{\mathbb{P}^{\theta,\dagger}_\tau[\hat{\bm{x}}]} \label{entropy-def},
\end{align}
the time-reversed path probability $\mathbb{P}^{\theta,\dagger}[\hat{\bm{x}}]$ generally involves four operations:
i) reversing the trajectory $\hat{\bm{x}} = (\bm{x}(t))_{t \in [0,\tau]} \rightarrow \hat{\bm{x}}^\dagger = (\bm{x}(\tau-t))_{t \in [0,\tau]}$, ii) reversing the protocol of any explicitly time-dependent parameters $\bm{a}(\bm{x},t) \rightarrow \bm{a}(\bm{x},\tau-t)$, iii) reversing the sign of odd-parity variables and iv) changing the discretization scheme from $\alpha \rightarrow 1-\alpha$.
Since in \eqref{langevin-mod}, there are no explicitly time-dependent parameters and all variables have even parity, the time-reversed path probability is given by
\begin{align}
\mathbb{P}^{\theta,\dagger}_K[\hat{\bm{x}}] = \prod_{k = 1}^K p_{dt,1}^\theta(\bm{x}_{K-k} \vert \bm{x}_{K-k+1}) p^\theta(\bm{x}_K,\tau).
\end{align}

\subsection{Stochastic entropy production}
Consequently, we may write the stochastic entropy production as
\begin{align}
\Sigma^\theta_\tau[\hat{\bm{x}}] = \lim_{dt \rightarrow 0} \sum_{k=1}^K \ln \frac{p_{dt,0}^\theta(\bm{x}_k \vert \bm{x}_{k-1})}{p_{dt,1}^\theta(\bm{x}_{k-1} \vert \bm{x}_{k})} + \ln \frac{p^\theta(\bm{x}(0),0)}{p^\theta(\bm{x}(\tau),\tau)}.
\end{align}
The second term is precisely the stochastic change in Shannon entropy.
In the first term, we have to evaluate
\begin{align}
\ln &\frac{p_{dt,0}^\theta(\bm{x}_k \vert \bm{x}_{k-1})}{p_{dt,1}^\theta(\bm{x}_{k-1} \vert \bm{x}_{k})} = - \frac{1}{4 dt} \bigg( \big( \bm{x}_k - \bm{x}_{k-1} - \bm{a}^\theta(\bm{x}_{k-1}) dt \big)^\text{T} \bm{B}^{-1}(\bm{x}_{k-1}) \big( \bm{x}_k - \bm{x}_{k-1} - \bm{a}^\theta(\bm{x}_{k-1}) dt \big) \\
& - \big[ \bm{x}_{k-1} - \bm{x}_{k} - \bm{a}^\theta(\bm{x}_{k-1}) dt - 2 \grad \bm{B}(\bm{x}_{k-1}) dt \big]^\text{T} \bm{B}^{-1}(\bm{x}_{k-1}) \big[ \bm{x}_{k-1} - \bm{x}_{k} - \bm{a}^\theta(\bm{x}_{k-1}) dt - 2 \grad \bm{B}(\bm{x}_{k-1}) dt \big] \bigg) \nn
&\qquad + \grad^\text{T} \bm{a}^\theta(\bm{x}_{k-1}) dt - \mathcal{H}_B(\bm{x}_{k-1}) \big) dt \n .
\end{align}
Canceling terms and using that $\bm{B}(\bm{x})$ is symmetric, we end up with
\begin{align}
\ln &\frac{p_{dt,0}^\theta(\bm{x}_k \vert \bm{x}_{k-1})}{p_{dt,1}^\theta(\bm{x}_{k-1} \vert \bm{x}_{k})} = \big[\bm{a}^{\theta}(\bm{x}_{k-1}) - \grad \bm{B}(\bm{x}_{k-1}) \big]^\text{T} \bm{B}^{-1}(\bm{x}_{k-1}) \big( \bm{x}_k - \bm{x}_{k-1} - \bm{a}^\theta(\bm{x}_{k-1}) dt \label{path-ratio} \\
& + \bigg(\big[ \bm{a}^{\theta}(\bm{x}_{k-1}) - \grad \bm{B}(\bm{x}_{k-1}) \big]^\text{T} \bm{B}^{-1}(\bm{x}_{k-1})  \big[ \bm{a}^{\theta}(\bm{x}_{k-1}) - \grad \bm{B}(\bm{x}_{k-1}) \big] + \grad^\text{T} \bm{a}^\theta(\bm{x}_{k-1}) - \mathcal{H}_B(\bm{x}_{k-1}) \bigg) dt \n .
\end{align}
In the continuous-time limit, this can be written as
\begin{align}
\Sigma^\theta_\tau[\hat{\bm{x}}] = \int_0^\tau dt \ \big[\bm{a}^\theta(\bm{x}(t)) - \grad \bm{B}(\bm{x}(t)) \big]^\text{T} \bm{B}^{-1}(\bm{x}(t)) \circ \dot{\bm{x}}(t) + \ln \frac{p^\theta(\bm{x}(0),0)}{p^\theta(\bm{x}(\tau),\tau)}, \label{path-entropy-0}
\end{align}
where $\circ$ is the Stratonovich product.
In order to show the equivalence of the two expressions, we define
\begin{align}
\bm{w}(\bm{x}) = \big[\bm{a}^\theta(\bm{x}) - \grad \bm{B}(\bm{x}) \big]^\text{T} \bm{B}^{-1}(\bm{x})
\end{align}
and use the Ito-formula
\begin{align}
 \bm{w}^\text{T}(\bm{x}(t)) \circ \dot{\bm{x}}(t) = \bm{w}^\text{T}(\bm{x}(t)) \cdot \dot{\bm{x}}(t) + \text{tr}\big(\bm{B}(\bm{x}(t)) \bm{\mathcal{J}}_w(\bm{x}(t)) \big),
\end{align}
where $\bm{\mathcal{J}}_w(\bm{x})$ is the Jacobian of $\bm{w}(\bm{x})$ and tr denotes the trace, to convert between the Stratonovich- and Ito-formulation.
Plugging this into \eqref{path-entropy-0}, \eqref{path-ratio} follows by explicit calculation.
We note that, in the steady state, the probability density is independent of $\theta$, $p_\text{st}^\theta(\bm{x}) = p_\text{st}(\bm{x})$.
To see this, we first consider the steady-state equation for $\theta = 1$,
\begin{align}
0 = - \grad^\text{T} \big(\bm{\nu}_\text{st}(\bm{x}) p_\text{st}(\bm{x}) \big) \qquad \text{with} \qquad \bm{\nu}_\text{st}(\bm{x}) = \bm{a}(\bm{x}) - \grad \bm{B}(\bm{x}) - \bm{B}(\bm{x}) \grad \ln p_\text{st}(\bm{x}) \label{steady-state} .
\end{align}
Writing the corresponding equation for arbitrary $\theta$ explicitly in terms of $\bm{\nu}_\text{st}(\bm{x})$, we have
\begin{align}
0 &= - \grad ^\text{T}\bigg( \theta \Big[ \bm{a}(\bm{x}) - \grad \bm{B}(\bm{x}) + \bm{B}(\bm{x})  \grad \ln p_\text{st}(\bm{x}) \Big] - \Big[ \grad \bm{B}(\bm{x}) + \bm{B}(\bm{x}) \grad \ln p_\text{st}(\bm{x}) \Big] - \grad \bm{B}(\bm{x}) \bigg) p^\theta_\text{st}(\bm{x}) \\
&= - \grad^\text{T} \bigg( \theta \bm{\nu}_\text{st}(\bm{x}) - \bm{B}(\bm{x}) \Big[\grad \ln \frac{p^\theta_\text{st}(\bm{x})}{p_\text{st}(\bm{x})} \Big] \bigg) p^\theta_\text{st}(\bm{x}) \n .
\end{align}
Comparing this to \eqref{steady-state}, it is obvious that $p_\text{st}^\theta(\bm{x}) = p_\text{st}(\bm{x})$ is a solution, and, since the solution to the present class of Fokker-Planck equations is unique, it is also the only one.
Further, we can identify the steady-state local mean velocity corresponding to \eqref{langevin-mod} as $\bm{\nu}^\theta_\text{st}(\bm{x}) = \theta \bm{\nu}_\text{st}(\bm{x})$.
In the steady state, we thus have for \eqref{path-entropy-0}
\begin{align}
\Sigma^\theta_\tau[\hat{\bm{x}}] = \int_0^\tau dt \ \big[\bm{a}^\theta(\bm{x}(t)) - \grad \bm{B}(\bm{x}(t)) \big]^\text{T} \bm{B}^{-1}(\bm{x}(t)) \circ \dot{\bm{x}}(t) + \ln \frac{p_\text{st}(\bm{x}(0))}{p_\text{st}(\bm{x}(\tau))}. \label{path-entropy}
\end{align}
Then, we may write the Shannon term as
\begin{align}
\ln \frac{p_\text{st}(\bm{x}(0))}{p_\text{st}(\bm{x}(\tau))} = - \int_0^\tau dt \ d_t \ln p_\text{st}(\bm{x}(t)) = - \int_0^\tau dt \ \grad \ln p(\bm{x}(t) \circ \dot{\bm{x}}(t).
\end{align}
Plugging this into \eqref{path-entropy}, we can combine the first and second term
\begin{align}
\Sigma^\theta_\tau[\hat{\bm{x}}] = \int_0^\tau dt \ {\underbrace{\big[\bm{a}^\theta(\bm{x}(t)) - \grad \bm{B}(\bm{x}(t)) - \bm{B}(\bm{x}(t)) \grad \ln p_\text{st}(\bm{x}(t)) \big]}_{= \bm{\nu}^\theta_\text{st}(\bm{x})}}^\text{T} \bm{B}^{-1}(\bm{x}(t)) \circ \dot{\bm{x}}(t) \label{path-entropy-meanvel} .
\end{align}
Using the scaling of the steady-state local mean velocity, we finally obtain
\begin{align}
\Sigma^\theta_\tau[\hat{\bm{x}}] = \theta \Sigma_\tau[\hat{\bm{x}}] = \theta \int_0^\tau dt \ \bm{\nu}^\text{T}_\text{st}(\bm{x}(t)) \bm{B}^{-1}(\bm{x}(t)) \circ \dot{\bm{x}}(t) \label{entropy-scaling} ,
\end{align}
which is \eqref{main-entropy-scaling} of the main text.
Summarizing the discussion so far, we have shown that choosing a weighting factor $\bm{\rho}(\bm{x}) = \bm{B}^{-1}(\bm{x}) \bm{\nu}_\text{st}(\bm{x})$ in the definition of the time-integrated current reproduces the stochastic entropy production and that \eqref{langevin-mod} leads to a factor $\theta$ in the latter compared to the dynamics at $\theta = 1$.

\subsection{Average entropy production}
Next, we want to evaluate the average of \eqref{entropy-def} with respect to the path probability $\mathbb{P}^\theta[\hat{\bm{x}}]$ in the steady state.
To do so, we first consider an arbitrary current
\begin{align}
J_\tau[\hat{\bm{x}}] &= \int_0^\tau dt \ \bm{w}^\text{T}(\bm{x}(t)) \circ \dot{\bm{x}}(t) = \int_0^\tau dt \ \Big(\bm{w}^\text{T}(\bm{x}(t)) \cdot \dot{\bm{x}}(t) + \text{tr}\big(\bm{B}(\bm{x}(t)) \bm{\mathcal{J}}_w(\bm{x}(t)) \big) \Big) \label{current} \\
&=\int_0^\tau dt \ \Big(\bm{w}^\text{T}(\bm{x}(t)) \big( \bm{a}^\theta(\bm{x}(t)) + \bm{G}(\bm{x}(t)) \cdot \bm{\xi}(t) \big) + \text{tr}\big(\bm{B}(\bm{x}(t)) \bm{\mathcal{J}}_w(\bm{x}(t)) \big) \Big) \n .
\end{align}
Since the noise is white, the average of the second term vanishes while the remaining terms only depend on the position at time $t$ and their average thus reduces to an average taken with respect to the steady-state probability density $p_\text{st}(\bm{x})$,
\begin{align}
\av{J_\tau}^\theta = \tau \Big( \Av{\bm{w}^\text{T} \bm{a}^\theta}_\text{st} + \Av{\text{tr}(\bm{B} \bm{\mathcal{J}}_w )}_\text{st} \Big).
\end{align}
In this expression, the second term is explicitly given by
\begin{align}
\Av{\text{tr}(\bm{B} \bm{\mathcal{J}}_w )}_\text{st} &= \int d\bm{x} \ p_\text{st}(\bm{x}) \sum_{i,j} B_{i j}(\bm{x}) \partial_{x_j} \rho_i(\bm{x}) \\
&=- \int d\bm{x} \ \sum_{i,j}  \rho_i(\bm{x}) \partial_{x_j} \big( B_{i j}(\bm{x}) p_\text{st}(\bm{x}) \big) \nn
&=- \int d\bm{x} \ \bm{w}^\text{T}(\bm{x}) \Big[ \grad \bm{B}(\bm{x}) + \bm{B}(\bm{x}) \grad \ln p_\text{st}(\bm{x}) \Big] p_\text{st}(\bm{x}) ,
\end{align}
where we integrated by parts from the first to the second line.
With this, we find
\begin{align}
\av{J_\tau}^\theta  = \tau \av{\bm{w}^\text{T} \bm{\nu}_\text{st}^\theta}_\text{st} = \tau \theta \av{\bm{w}^\text{T} \bm{\nu}_\text{st}}_\text{st} = \theta \av{J_\tau} \label{current-average} .
\end{align}
Applying this result to the average of $\Sigma_\tau^\theta$, \eqref{path-entropy}, we obtain
\begin{align}
\Av{\Sigma_\tau^\theta}^\theta = \theta^2 \tau \int d\bm{x} \ \bm{\nu}_\text{st}^\text{T}(\bm{x}) \bm{B}^{-1}(\bm{x}) \bm{\nu}_\text{st}(\bm{x}) = \theta^2 \Av{\Sigma_\tau} = \theta^2 \Delta S_{\text{irr},\tau}.
\end{align}
We remark that since $\bm{B}(\bm{x})$ is positive definite, this expression is always positive, except when the local mean velocity vanishes.
The positivity also follows from the fact that the average of $\Sigma_\tau^\theta$ may be expressed as a Kullback-Leibler (KL) divergence,
\begin{align}
\Av{\Sigma_\tau^\theta}^\theta = \int \mathcal{D}\hat{\bm{x}} \ \mathbb{P}_\tau^\theta[\hat{\bm{x}}] \ln \frac{\mathbb{P}_\tau^\theta[\hat{\bm{x}}]}{\mathbb{P}_\tau^{\theta,\dagger}[\hat{\bm{x}}]} = D_\text{KL} \big(\mathbb{P}_\tau^\theta[\hat{\bm{x}}] \hspace{.25mm} \Vert \hspace{.25mm} \mathbb{P}_\tau^{\theta,\dagger}[\hat{\bm{x}}] \big).
\end{align}

\subsection{Different values of $\theta$}
While in \eqref{entropy-def}, we computed the log-ratio between the forward and time-reverse path probability density for a single value of $\theta$, we may formally also choose different values of $\theta$ and compute the log-ratios between the forward, and the forward and time-reverse path probabilities,
\begin{subequations}
\begin{align}
\Xi^{\theta_1,\theta_2}_\tau[\hat{\bm{x}}] &= \ln \frac{\mathbb{P}_\tau^{\theta_1}[\hat{\bm{x}}]}{\mathbb{P}_\tau^{\theta_2}[\hat{\bm{x}}]} \label{forward-forward} \\
\tilde{\Xi}^{\theta_1,\theta_2}_\tau[\hat{\bm{x}}] &= \ln \frac{\mathbb{P}_\tau^{\theta_1}[\hat{\bm{x}}]}{\mathbb{P}_\tau^{\theta_2,\dagger}[\hat{\bm{x}}]} \label{forward-backward} .
\end{align}
\end{subequations}
\eqref{forward-forward} is the path ratio between two path probabilities with different drift vectors.
This is given by Girsanov's lemma \cite{Gir60,Dec17}, and its average evaluates to
\begin{align}
\Av{\Xi^{\theta_1,\theta_2}_\tau}^{\theta_1} &= D_\text{KL} \big(\mathbb{P}_\tau^{\theta_1}[\hat{\bm{x}}] \hspace{.25mm} \Vert \hspace{.25mm} \mathbb{P}_\tau^{\theta_2}[\hat{\bm{x}}] \big) = \frac{\tau}{4} \int d\bm{x} \ \big(\bm{a}^{\theta_1}(\bm{x}) - \bm{a}^{\theta_2}(\bm{x}) \big)^\text{T} \bm{B}^{-1}(\bm{x}) \big(\bm{a}^{\theta_1}(\bm{x}) - \bm{a}^{\theta_2}(\bm{x}) \big) p_\text{st}(\bm{x}) \label{forward-forward-1} \\
&=\frac{\tau}{4} \big(\theta_1 - \theta_2)^2 \int d\bm{x} \ \bm{\nu}_\text{st}^\text{T}(\bm{x}) \bm{B}^{-1}(\bm{x}) \bm{\nu}_\text{st}(\bm{x}) p_\text{st}(\bm{x}) \nn
&= \frac{1}{4} \big(\theta_1 - \theta_2)^2 \Delta S^\text{irr}_\tau \n .
\end{align}
The KL divergence is a measure of how distinguishable two probability densities are, and thus the entropy production acts as a distance measure on the family of dynamics \eqref{langevin-mod}.
Next, for \eqref{forward-backward} we find, in analogy to \eqref{path-ratio}
\begin{align}
\ln &\frac{p_{dt,0}^{\theta_1}(\bm{x}_k \vert \bm{x}_{k-1})}{p_{dt,1}^{\theta_2}(\bm{x}_{k-1} \vert \bm{x}_{k})} = \frac{1}{2} \big[\bm{a}^{\theta_1}(\bm{x}_{k-1}) + \bm{a}^{\theta_2}(\bm{x}_{k-1}) - 2 \grad \bm{B}(\bm{x}_{k-1}) \big]^\text{T} \bm{B}^{-1}(\bm{x}_{k-1})  \big( \bm{x}_k - \bm{x}_{k-1} - \bm{a}^{\theta_1}(\bm{x}_{k-1}) dt \big) \\
&+ \frac{1}{4} \big[\bm{a}^{\theta_1}(\bm{x}_{k-1}) + \bm{a}^{\theta_2}(\bm{x}_{k-1}) - 2 \grad \bm{B}(\bm{x}_{k-1}) \big]^\text{T} \bm{B}^{-1}(\bm{x}_{k-1}) \big[\bm{a}^{\theta_1}(\bm{x}_{k-1}) + \bm{a}^{\theta_2}(\bm{x}_{k-1}) - 2 \grad \bm{B}(\bm{x}_{k-1})\big] dt \nn
&+ \grad \bm{a}^{\theta_2}(\bm{x}_{k-1}) dt - \mathcal{H}_B(\bm{x}_{k-1}) dt \n .
\end{align}
Taking the average, the first term again cancels, and we have
\begin{align}
&\Av{\tilde{\Xi}^{\theta_1,\theta_2}_\tau}^{\theta_1} = D_\text{KL} \big(\mathbb{P}_\tau^{\theta_1}[\hat{\bm{x}}] \hspace{.25mm} \Vert \hspace{.25mm} \mathbb{P}_\tau^{\theta_2,\dagger}[\hat{\bm{x}}] \big) \\
&\quad = \tau \int d\bm{x} \ \bigg[ \frac{1}{4} \Big[\bm{a}^{\theta_1}(\bm{x}) + \bm{a}^{\theta_2}(\bm{x}) - 2 \grad \bm{B}(\bm{x}) \Big]^\text{T} \bm{B}^{-1}(\bm{x}) \Big[\bm{a}^{\theta_1}(\bm{x}) + \bm{a}^{\theta_2}(\bm{x}) - 2 \grad \bm{B}(\bm{x})\Big] + \grad^\text{T} \bm{a}^{\theta_2}(\bm{x}) - \mathcal{H}_B(\bm{x}) \bigg] p_\text{st}(\bm{x}) \n .
\end{align}
Using the explicit expression for $\bm{a}^\theta(\bm{x})$, we rewrite
\begin{align}
\bm{a}^{\theta_1}(\bm{x}) + \bm{a}^{\theta_2}(\bm{x}) - 2 \grad \bm{B}(\bm{x}) = \big(\theta_1 + \theta_2 \big) \bm{\nu}_\text{st}(\bm{x}) + 2 \bm{B}(\bm{x}) \grad \ln p_\text{st}(\bm{x}) .
\end{align}
Then we obtain
\begin{align}
\Av{\Xi^{\theta_1,\theta_2}_\tau}^{\theta_1} &= \tau \int d\bm{x} \ \bigg[ \frac{1}{4} \big(\theta_1 + \theta_2\big)^2 \bm{\nu}_\text{st}^\text{T}  \bm{B}^{-1}(\bm{x}) \bm{\nu}_\text{st} + \big(\theta_1 + \theta_2 \big) \bm{\nu}_\text{st}^\text{T}(\bm{x}) \grad \ln p_\text{st}(\bm{x}) + \big[ \grad \ln p_\text{st}(\bm{x}) \big]^\text{T} \bm{B}(\bm{x}) \big[ \grad \ln p_\text{st}(\bm{x}) \big]  \nn
&\hspace{2cm} + \theta_2 \grad^\text{T} \bm{\nu}_\text{st}(\bm{x}) + \grad^\text{T} \Big( \grad \bm{B}(\bm{x}) + \bm{B}(\bm{x}) \grad \ln p_\text{st}(\bm{x}) \Big) - \mathcal{H}_B(\bm{x}) \bigg] p_\text{st}(\bm{x}) .
\end{align}
Using \eqref{steady-state}, we have $\grad^\text{T} \bm{\nu}_\text{st}(\bm{x}) = -\bm{\nu}_\text{st}^\text{T}(\bm{x}) \grad \ln p_\text{st}(\bm{x})$.
Further, we have that $\mathcal{H}_B(\bm{x}) = \grad^\text{T} (\grad \bm{B}(\bm{x}))$ and thus
\begin{align}
\Av{\tilde{\Xi}^{\theta_1,\theta_2}_\tau}^{\theta_1} &= \tau \int d\bm{x} \ \bigg[ \frac{1}{4} \big(\theta_1 + \theta_2\big)^2 \bm{\nu}_\text{st}^\text{T}  \bm{B}^{-1}(\bm{x}) \bm{\nu}_\text{st} + \theta_1 \bm{\nu}_\text{st}^\text{T}(\bm{x}) \grad \ln p_\text{st}(\bm{x}) + \big[ \grad \ln p_\text{st}(\bm{x}) \big]^\text{T} \bm{B}(\bm{x}) \big[ \grad \ln p_\text{st}(\bm{x}) \big]  \nn
&\hspace{2cm} + \grad^\text{T} \big(\bm{B}(\bm{x}) \grad \ln p_\text{st}(\bm{x}) \big) \bigg] p_\text{st}(\bm{x}) \label{forward-backward-1} .
\end{align}
Next, we are going to prove the identity
\begin{align}
I \equiv \int d\bm{x} \ p_\text{st}(\bm{x}) \bigg( \big[\grad \ln p_\text{st}(\bm{x}) \big]^\text{T} \bm{B}(\bm{x}) \big[\grad \ln p_\text{st}(\bm{x}) \big] + \grad^\text{T} \bm{a}(\bm{x}) - \mathcal{H}_B(\bm{x}) \bigg) = 0 \label{identity-1}.
\end{align}
To do so, we introduce the function $\psi(\bm{x}) = -\ln p_\text{st}(\bm{x})$, for which we obtain from \eqref{steady-state} the nonlinear partial differential equation
\begin{align}
0 = \bm{a}^\text{T}(\bm{x}) \grad \psi(\bm{x}) - \grad^\text{T} \bm{a}(\bm{x}) - \grad^\text{T} \big(\grad \bm{B}(\bm{x}) \psi(\bm{x}) \big) + \big[\grad \psi(\bm{x}) \big]^\text{T} \bm{B}(\bm{x}) \big[\grad \psi(\bm{x}) \big] + \big(1+\psi(\bm{x}) \big) \mathcal{H}_B(\bm{x}) .
\end{align}
Using this, we can write the left-hand side of \eqref{identity-1} as
\begin{align}
I = 2 \int d\bm{x} \ p_\text{st}(\bm{x}) \bigg( \grad^\text{T} \bm{a}(\bm{x}) - \mathcal{H}_B(\bm{x}) \bigg) \underbrace{- \int d\bm{x} \ p_\text{st}(\bm{x}) \bigg( \bm{a}^\text{T}(\bm{x}) \grad \psi(\bm{x}) - \grad^\text{T} \big(\grad \bm{B}(\bm{x}) \psi(\bm{x}) \big) + \psi(\bm{x}) \mathcal{H}_B(\bm{x}) \bigg)}_{I_1} \n .
\end{align}
In order to facilitate the following calculation, we write the second term explicitly in component form
\begin{align}
I_1 = - \int d\bm{x} \ p_\text{st}(\bm{x}) \bigg( \sum_i a_i(\bm{x}) \partial_{x_i} \psi(\bm{x}) - \sum_{i, j}\partial_{x_i} \partial_{x_j} \big( B_{i j}(\bm{x}) \psi(\bm{x}) \big) + \psi(\bm{x}) \sum_{i, j}\partial_{x_i} \partial_{x_j} B_{i j}(\bm{x}) \bigg) .
\end{align}
We integrate by parts with respect to $x_i$ in the first term,
\begin{align}
I_1 = \int d\bm{x} \  \bigg( \psi(\bm{x}) \sum_i \partial_{x_i} \big( a_i(\bm{x}) p_\text{st}(\bm{x}) \big) + p_\text{st}(\bm{x}) \sum_{i, j}\partial_{x_i} \partial_{x_j} \big( B_{i j}(\bm{x}) \psi(\bm{x}) \big)  - p_\text{st}(\bm{x}) \psi(\bm{x}) \sum_{i, j} \partial_{x_i} \partial_{x_j} B_{i j}(\bm{x}) \bigg).
\end{align}
Once again, we use \eqref{steady-state} to write
\begin{align}
\sum_i \partial_{x_i} \big( a_i(\bm{x}) p_\text{st}(\bm{x}) \big) = \sum_{i,j} \partial_{x_i} \partial_{x_j} \big( B_{i j}(\bm{x}) p_\text{st}(\bm{x}) \big)
\end{align}
and, using this,
\begin{align}
I_1 = \sum_{i,j} \int d\bm{x} \  \bigg( \psi(\bm{x}) \partial_{x_i} \partial_{x_j} \big( B_{i j}(\bm{x}) p_\text{st}(\bm{x}) \big) + p_\text{st}(\bm{x}) \partial_{x_i} \partial_{x_j} \big( B_{i j}(\bm{x}) \psi(\bm{x}) \big)  - p_\text{st}(\bm{x}) \psi(\bm{x}) \partial_{x_i} \partial_{x_j} B_{i j}(\bm{x}) \bigg).
\end{align}
We again integrate by parts with respect to $x_i$ and use $\partial_{x_i} p_\text{st}(\bm{x}) = - p_\text{st}(\bm{x}) \partial_{x_i} \psi(\bm{x})$ to write
\begin{align}
I_1 &= 2 \sum_{i,j} \int d\bm{x} \ p_\text{st}(\bm{x}) \big[\partial_{x_i}\psi(\bm{x}) \big] B_{i j}(\bm{x}) \big[\partial_{x_j}\psi(\bm{x}) \big] \\
& \qquad - \sum_{i,j} \int d\bm{x} \ \bigg( - \big[\partial_{x_i} p_\text{st}(\bm{x})  \big] \partial_{x_j} B_{i j}(\bm{x}) + \psi(\bm{x}) \big[\partial_{x_i} p_\text{st}(\bm{x}) \big] \partial_{x_j} B_{i j}(\bm{x}) - \big[\partial_{x_i} \big( p_\text{st}(\bm{x})  \psi(\bm{x}) \big) \big] \partial_{x_j} B_{i j}(\bm{x}) \bigg) \nn
&= 2 \int d\bm{x} \ p_\text{st} \big[\grad \psi(\bm{x})\big]^\text{T} \bm{B}(\bm{x}) \big[\grad \psi(\bm{x})\big] , \n
\end{align}
since the terms under the second integral cancel.
We thus obtain for the left-hand side of \eqref{identity-1},
\begin{align}
I = 2 I,
\end{align}
which implies $I = 0$ and thus \eqref{identity-1}.
This has two consequences:
First, integrating by parts, we can rewrite $I$ as
\begin{align}
I = \int d\bm{x} \ \bm{\nu}_\text{st}^\text{T}(\bm{x}) \grad p_\text{st}(\bm{x}) = - \int d\bm{x} \ p_\text{st}(\bm{x}) \grad^\text{T} \bm{\nu}_\text{st}(\bm{x}) = 0 .
\end{align}
Applying this to \eqref{forward-backward-1}, we see that the term proportional to $\theta_1$ vanishes,
\begin{align}
\Av{\Xi^{\theta_1,\theta_2}_\tau}^{\theta_1} &= \tau \int d\bm{x} \ \bigg[ \frac{1}{4} \big(\theta_1 + \theta_2\big)^2 \bm{\nu}_\text{st}^\text{T}  \bm{B}^{-1}(\bm{x}) \bm{\nu}_\text{st} + \big[ \grad \ln p_\text{st}(\bm{x}) \big]^\text{T} \bm{B}(\bm{x}) \big[ \grad \ln p_\text{st}(\bm{x}) \big]  \nn
&\hspace{2cm} + \grad^\text{T} \big(\bm{B}(\bm{x}) \grad \ln p_\text{st}(\bm{x}) \big) \bigg] p_\text{st}(\bm{x}) .
\end{align}
Second, we can replace the second term in the above expression using \eqref{identity-1},
\begin{align}
\Av{\tilde{\Xi}^{\theta_1,\theta_2}_\tau}^{\theta_1} &= \tau \int d\bm{x} \ \bigg[ \frac{1}{4} \big(\theta_1 + \theta_2\big)^2 \bm{\nu}_\text{st}^\text{T}  \bm{B}^{-1}(\bm{x}) \bm{\nu}_\text{st} - \grad^\text{T} \bm{\nu}_\text{st}(\bm{x}) \bigg] p_\text{st}(\bm{x}) ,
\end{align}
where, again the second term cancels.
Finally, we then find for the path ratio \eqref{forward-backward}
\begin{align}
\Av{\tilde{\Xi}^{\theta_1,\theta_2}_\tau}^{\theta_1} = D_\text{KL} \big(\mathbb{P}_\tau^{\theta_1}[\hat{\bm{x}}] \hspace{.25mm} \Vert \hspace{.25mm} \mathbb{P}_\tau^{\theta_2,\dagger}[\hat{\bm{x}}] \big) = \frac{1}{4} \big(\theta_1 + \theta_2 \big)^2 \Delta S^\text{irr}_\tau.
\end{align}
In particular, for $\theta_1 = -\theta$ and $\theta_2 = \theta$, this gives
\begin{align}
D_\text{KL} \big(\mathbb{P}_\tau^{-\theta}[\hat{\bm{x}}] \hspace{.25mm} \Vert \hspace{.25mm} \mathbb{P}_\tau^{\theta,\dagger}[\hat{\bm{x}}] \big) = 0 \label{KL-time-reverse} ,
\end{align}
which is \eqref{main-KL-time-reverse} of the main text.
This implies that the time-reversed path probability density for $\theta$ is indistinguishable from the forward path probability density for $-\theta$, and thus, that the dynamics at $-\theta$ is equivalent to the time-reversed dynamics at $\theta$.

\section{Fluctuations of stochastic currents}
Here, we want to obtain an explicit expression for the variance of a time-integrated stochastic current defined in \eqref{current}.
We broadly follow the calculation of Ref.~\cite{Pig17}.
Formally, we may write the second moment of the current as the average
\begin{align}
\Av{\big(J_\tau\big)^2}^\theta = \int_0^\tau dt \int_0^\tau ds \ \Av{\Big(\bm{w}^\text{T}(t) \cdot \dot{\bm{x}}(t) + \text{tr}\big(\bm{B}(t) \bm{\mathcal{J}}_w(t) \big) \Big) \Big(\bm{w}^\text{T}(s) \cdot \dot{\bm{x}}(s) + \text{tr}\big(\bm{B}(s) \bm{\mathcal{J}}_w(s) \big) \Big)}^\theta ,
\end{align}
where, in the interest of a more compact notation, we write $f(\bm{x}(t)) = f(t)$.
First, we note that the average is symmetric with respect to exchanging $t$ and $s$ and thus,
\begin{align}
\Av{\big(J_\tau\big)^2}^\theta = 2 \int_0^\tau dt \int_0^t ds \ \Av{\Big(\bm{w}^\text{T}(t) \cdot \dot{\bm{x}}(t) + \text{tr}\big(\bm{B}(t) \bm{\mathcal{J}}_w(t) \big) \Big) \Big(\bm{w}^\text{T}(s) \cdot \dot{\bm{x}}(s) + \text{tr}\big(\bm{B}(s) \bm{\mathcal{J}}_w(s) \big) \Big)}^\theta .
\end{align}
Replacing $\dot{\bm{x}}(t)$ using \eqref{langevin-mod}, we find that there are three distinct contributions
\begin{align}
\Av{\big(J_\tau\big)^2}^\theta &= 2 \int_0^\tau dt \int_0^t ds \ \Av{\Big(\bm{w}^\text{T}(t) \bm{a}^\theta(t) + \text{tr}\big(\bm{B}(t) \bm{\mathcal{J}}_w(t) \big) \Big) \Big(\bm{w}^\text{T}(s) \bm{a}^\theta(s) + \text{tr}\big(\bm{B}(s) \bm{\mathcal{J}}_w(s) \big) \Big)}^\theta \\
&\qquad+ 2\int_0^\tau dt \int_0^t ds \ \Av{\Big(\bm{w}^\text{T}(t) \bm{a}^\theta(t) + \text{tr}\big(\bm{B}(t) \bm{\mathcal{J}}_w(t) \big) \Big) \Big(\bm{w}^\text{T}(s) \bm{G}(s) \cdot \bm{\xi}(s) \Big) + \Big[ t \leftrightarrow s \Big]}^\theta \nn
&\qquad+\int_0^\tau dt \int_0^\tau ds \ \Av{\Big(\bm{w}^\text{T}(t) \bm{G}(t) \cdot \bm{\xi}(t) \Big) \Big(\bm{w}^\text{T}(s) \bm{G}(s) \cdot \bm{\xi}(s) \Big)}^\theta . \n
\end{align}
We now introduce the local mean velocity by writing (see \eqref{langevin-mod})
\begin{align}
\bm{a}^\theta(\bm{x}) = \theta \bm{\nu}_\text{st}(\bm{x}) + \bm{\phi}_\text{st}(\bm{x}),
\end{align}
and define the functions
\begin{align}
\mu(\bm{x}) = \bm{w}^\text{T}(\bm{x}) \bm{\nu}_\text{st}(\bm{x}), \qquad \psi(\bm{x}) = \bm{w}^\text{T}(\bm{x}) \bm{\phi}_\text{st}(\bm{x}) + \text{tr}\big(\bm{B}(\bm{x}) \bm{\mathcal{J}}_w(\bm{x}) \big) \label{mu-psi} .
\end{align}
In terms of this notation, we can write the second moment as
\begin{align}
\Av{\big(J_\tau\big)^2}^\theta &= 2\int_0^\tau dt \int_0^t ds \ \Av{\Big( \theta \mu(t) + \psi(t) \Big) \Big( \theta \mu(s) + \psi(s) \Big)}^\theta \label{current-variance-1} \\
&\qquad+2\int_0^\tau dt \int_0^t ds \ \Av{\Big(\theta \mu(t) + \psi(t) \Big) \Big(\bm{w}^\text{T}(s) \bm{G}(s) \cdot \bm{\xi}(s) \Big) + \Big(\theta \mu(s) + \psi(s) \Big) \Big(\bm{w}^\text{T}(t) \bm{G}(t) \cdot \bm{\xi}(t) \Big)}^\theta \nn
&\qquad+\int_0^\tau dt \int_0^\tau ds \ \Av{\Big(\bm{w}^\text{T}(t) \bm{G}(t) \cdot \bm{\xi}(t) \Big) \Big(\bm{w}^\text{T}(s) \bm{G}(s) \cdot \bm{\xi}(s) \Big)}^\theta . \n
\end{align}
Since the noise is white, the term in the third line only contributes for $t = s$ and evaluates to
\begin{align}
\int_0^\tau dt \int_0^\tau ds \ \Av{\Big(\bm{w}^\text{T}(t) \bm{G}(t) \cdot \bm{\xi}(t) \Big) \Big(\bm{w}^\text{T}(s) \bm{G}(s) \cdot \bm{\xi}(s) \Big)}^\theta = 2 \int_0^\tau dt \ \Av{\bm{w}^\text{T}(t) \bm{B}(t) \bm{w}(t)}^\theta,
\end{align}
where we used $\bm{G}(\bm{x}) \bm{G}^\text{T}(\bm{x}) = 2 \bm{B}(\bm{x})$ and the noise correlation ($\bm{1}$ is the $N\times N$ identity matrix)
\begin{align}
\av{\bm{\xi}^\text{T}(t) \bm{\xi}(s)} = \bm{1} \delta(t-s).
\end{align}
Likewise, the white-noise property means that $\bm{\xi}(s)$ is independent of the trajectory $\bm{x}(t)$ for $t < s$, and thus the second term in the second line vanishes.
However, since the trajectory for $t > s$ depends on the value of the noise at time $s$, we need to evaluate these correlations.
A convenient way to do this is via Doob-conditioning \cite{Doo57,Che15}, where a stochastic process $\bm{z}(s)$, $s < t$ is constructed, which corresponds to the subset of trajectories of \eqref{langevin-mod}, which end at a fixed value $\bm{x}(t) = \bm{x}_0$.
This method has been used in Ref.~\cite{Pig17} to evaluate the fluctuations of the entropy production.
As a result, we find a general expression for a correlation function of the type
\begin{align}
\Av{f(\bm{x}(t),\bm{x}(s)) \cdot \bm{\xi}(s)} = \left\lbrace \begin{array}{ll}
\int d\bm{x} \int d\bm{y} \ p(\bm{x},t;\bm{y},s) f(\bm{x},\bm{y}) \bm{G}^\text{T}(\bm{y}) \grad_y \ln p(\bm{x},t \vert \bm{y},s) &\text{for} \quad t > s \\[1ex]
0 &\text{for} \quad t \leq s,
\end{array} \right.
\end{align}
where $f(\bm{x},\bm{y})$ is a differentiable function and $p(\bm{x},t ; \bm{y}, s)$ denotes the joint and $p(\bm{x},t \vert \bm{y},s) = p(\bm{x},t ; \bm{y}, s)/p(\bm{y},s)$ the conditional probability density.
Applying this to the above expression, we find
\begin{align}
\int_0^\tau &dt \int_0^t ds \ \Av{\Big( \theta \mu(t) + \psi(t) \Big) \Big(\bm{w}^\text{T}(s) \bm{G}(s) \cdot \bm{\xi}(s) \Big)}^\theta \\
&= \int_0^\tau dt \int_0^t ds \int d\bm{x} \int d\bm{y} \ p^\theta(\bm{x},t ; \bm{y},s) \Big(\theta \mu(\bm{x}) + \psi(\bm{x}) \Big) \Big(\bm{w}^\text{T}(\bm{y}) \bm{B}(\bm{y}) \grad_y \ln p^\theta(\bm{x},t \vert \bm{y},s) \Big) \n .
\end{align}
To proceed, we focus on the steady state.
Then, integrating by parts with respect to $\bm{y}$ and using $p^\theta(\bm{x},t;\bm{y},s) = p^\theta(\bm{x},t \vert \bm{y},s) p_\text{st}(\bm{y})$, we obtain
\begin{align}
\int_0^\tau &dt \int_0^t ds \ \Av{\Big( \theta \mu(t) + \psi(t) \Big) \Big(\bm{w}^\text{T}(s) \bm{G}(s) \cdot \bm{\xi}(s) \Big)}^\theta \label{second-term} \\
&= - \int_0^\tau dt \int_0^t ds \int d\bm{x} \int d\bm{y} \ p^\theta(\bm{x},t ; \bm{y},s) \Big(\theta \mu(\bm{x}) + \psi(\bm{x}) \Big) \psi(\bm{y}) \n .
\end{align}
Summing up, we obtain for the variance of the current
\begin{align}
\text{Var}^\theta(J_\tau) &= 2\int_0^\tau dt \int_0^t ds \int d\bm{x} \int d\bm{y} \ \Big( \theta \mu(\bm{x}) + \psi(\bm{x}) \Big) \Big( \theta \mu(\bm{y}) + \psi(\bm{y}) \Big) p^\theta(\bm{x},t ; \bm{y},s) \\
&\quad - 4\int_0^\tau dt \int_0^t ds \int d\bm{x} \int d\bm{y} \ \Big( \theta \mu(\bm{x}) + \psi(\bm{x}) \Big) \psi(\bm{y}) p^\theta(\bm{x},t ; \bm{y},s) \nn
&\quad + 2 \tau \int d\bm{x} \ \bm{\rho}^\text{T}(\bm{x}) \bm{B}(\bm{x}) \bm{\rho}(\bm{x}) p_\text{st}(\bm{x}) - \big(\av{r_\tau}^\theta \big)^2 \n .
\end{align}
Canceling terms and using the result for the average \eqref{current-average}, we finally find
\begin{align}
\text{Var}^\theta(J_\tau) &= 2 \theta^2 \int_0^\tau dt \int_0^t ds \int d\bm{x} \int d\bm{y} \ \mu(\bm{x}) \mu(\bm{y}) \Big( p^\theta(\bm{x},t ; \bm{y},s) - p_\text{st}(\bm{x}) p_\text{st}(\bm{y}) \Big) \label{current-variance} \\
&\quad + 2\theta \int_0^\tau dt \int_0^t ds \int d\bm{x} \int d\bm{y} \ \Big(\mu(\bm{y}) \psi(\bm{x}) - \mu(\bm{x}) \psi(\bm{y}) \Big) p^\theta(\bm{x},t ; \bm{y},s) \nn
&\quad + 2 \tau \int d\bm{x} \ \bm{w}^\text{T}(\bm{x}) \bm{B}(\bm{x}) \bm{w}(\bm{x}) p_\text{st}(\bm{x}) - 2 \int_0^\tau dt \int_0^t ds \int d\bm{x} \int d\bm{y} \ \psi(\bm{x}) \psi(\bm{y}) p^\theta(\bm{x},t ; \bm{y},s) \n .
\end{align}
This can be related to the local mean current defined in \eqref{main-current-splitting},
\begin{align}
\bar{J}^\theta_\tau[\hat{\bm{x}}] = \int_0^\tau dt \ \bm{w}^\text{T}(\bm{x}) \bm{\nu}^\theta_\text{st}(\bm{x}),
\end{align}
by noting that the term in the first line is precisely the variance of this quantity
\begin{align}
\text{Var}^\theta(\bar{J}^\theta_\tau) = 2 \theta^2 \int_0^\tau dt \int_0^t ds \int d\bm{x} \int d\bm{y} \ \mu(\bm{x}) \mu(\bm{y}) \Big( p^\theta(\bm{x},t ; \bm{y},s) - p_\text{st}(\bm{x}) p_\text{st}(\bm{y}) \Big) .
\end{align}
Similarly, the terms in the third line of \eqref{current-variance} are obtained as the variance of the relative current $\delta J^\theta_\tau = J_\tau - \bar{J}^\theta_\tau$,
\begin{align}
\text{Var}^\theta(\delta J^\theta_\tau) = 2 \tau \int d\bm{x} \ \bm{\rho}^\text{T}(\bm{x}) \bm{B}(\bm{x}) \bm{w}(\bm{x}) p_\text{st}(\bm{x}) - 2 \int_0^\tau dt \int_0^t ds \int d\bm{x} \int d\bm{y} \ \psi(\bm{x}) \psi(\bm{y}) p^\theta(\bm{x},t ; \bm{y},s) .
\end{align}
Then, the term in the second line of \eqref{current-variance} is immediately identified as the covariance,
\begin{align}
\text{Cov}^\theta(\bar{J}^\theta_\tau,\delta J^\theta_\tau) = \theta \int_0^\tau dt \int_0^t ds \int d\bm{x} \int d\bm{y} \ \Big(\mu(\bm{y}) \psi(\bm{x}) - \mu(\bm{x}) \psi(\bm{y}) \Big) p^\theta(\bm{x},t ; \bm{y},s) .
\end{align}

For the specific observable $J_\tau = \Sigma^\theta_\tau$, i.~e.~the stochastic entropy production \eqref{entropy-def}, we saw in \eqref{path-entropy-meanvel} that the corresponding weighting function is given by $\bm{w}(\bm{x}) = \bm{B}^{-1}(\bm{x}) \bm{\nu}^\theta_\text{st}(\bm{x})$.
We calculate the function $\psi(\bm{x})$ (see \eqref{mu-psi}) for this choice of $\bm{\rho}(\bm{x})$,
\begin{align}
\psi(\bm{x}) = \bm{\nu}^{\theta,\text{T}}_\text{st}(\bm{x}) \bm{B}^{-1}(\bm{x}) \big(\grad^\text{T} \bm{B}(\bm{x}) \big) + \bm{\nu}^{\theta,\text{T}}_\text{st}(\bm{x})  \grad \ln p_\text{st}(\bm{x}) + \text{tr} \big(\bm{C}(\bm{x}) \big),
\end{align}
where the matrix $\bm{C}(\bm{x})$ has entries
\begin{align}
C_{i j}(\bm{x}) &= \sum_{k,l} B_{i k}(\bm{x}) \partial_{x_j} \Big( \nu_{\text{st},l}^\theta(\bm{x}) \big(\bm{B}^{-1}(\bm{x}) \big)_{l k} \Big) \\
&= \partial_{x_j} \nu_{\text{st},j}^\theta(\bm{x}) - \sum_{k,l} B_{i k}(\bm{x}) \nu_{\text{st},l}^\theta(\bm{x})   \big(\bm{B}^{-1}(\bm{x}) [\partial_{x_j} \bm{B}(\bm{x})] \bm{B}^{-1}(\bm{x})\big)_{l k} \n .
\end{align}
The second term cancels the first term in the above expression for $\psi(\bm{x})$ and we obtain
\begin{align}
\psi(\bm{x}) = \grad^\text{T} \bm{\nu}_\text{st}^\theta(\bm{x}) + \bm{\nu}^{\theta,\text{T}}_\text{st}(\bm{x})  \grad \ln p_\text{st}(\bm{x}) = \frac{1}{ p_\text{st}(\bm{x})} \grad^\text{T} \big( \bm{\nu}_\text{st}^\theta(\bm{x})  p_\text{st}(\bm{x}) \big) = 0,
\end{align}
since this is precisely the steady state condition.
For the stochastic entropy production, we thus find
\begin{align}
\text{Var}^\theta(\delta \Sigma_\tau^\theta) &= 2 \tau \int d\bm{x} \ \bm{\nu}^{\theta,\text{T}}_\text{st}(\bm{x}) \bm{B}^{-1}(\bm{x}) \bm{\nu}^{\theta}_\text{st}(\bm{x}) = 2 \Delta S^{\text{irr},\theta}_\tau \qquad \text{and} \qquad \text{Cov}^\theta(\bar{\Sigma}_\tau^\theta, \delta \Sigma_\tau^\theta) = 0 \label{entropy-fluctuations}. 
\end{align}
This means that $\bar{\Sigma}_\tau$ and $\delta \Sigma_\tau$ are statistically independent and the variance of the latter is precisely twice the average entropy production.
We thus obtain \eqref{main-entropy-splitting} of the main text for $\theta = 1$,
\begin{align}
\text{Var}(\Sigma_\tau) = \text{Var}(\bar{\Sigma}_\tau) + \text{Var}(\delta \Sigma_\tau) \label{entropy-splitting} .
\end{align}
In Ref.~\cite{Pig17}, the first term was identified with the fluctuations of an entropic time.
Further, we find from \eqref{current-variance} for the variance of a current in the equilibrium dynamics at $\theta = 0$
\begin{align}
\text{Var}^0(J_\tau) &= 2 \tau \int d\bm{x} \ \bm{w}^\text{T}(\bm{x}) \bm{B}(\bm{x}) \bm{w}(\bm{x}) p_\text{st}(\bm{x}) - 2 \int_0^\tau dt \int_0^t ds \int d\bm{x} \int d\bm{y} \ \psi(\bm{x}) \psi(\bm{y}) p^0(\bm{x},t ; \bm{y},s) .
\end{align}
For the stochastic entropy production, we saw above that we have $\psi(\bm{x}) = 0$ and thus
\begin{align}
\text{Var}^0(\Sigma_\tau) = 2 \Delta S^\text{irr}_\tau ,
\end{align}
which, together with \eqref{entropy-fluctuations} gives \eqref{main-entropy-fluctuations} of the main text.

\section{Bounds on the current cumulant generating function and comparison between continuous and discrete time-reversal}
Using the continuous time-reversal and in particular \eqref{forward-forward-1}, we can give a concise derivation of the quadratic lower bound on the cumulant generating function discussed in Refs.~\cite{Nem11,Gin16,Pie16}.
We start from the Kullback-inequality \cite{Kul54}
\begin{align}
D_\text{KL}(p^b \Vert p^a) \geq \sup_{h} \Big( h \av{r}^b - K_r^a(h) \Big),
\end{align}
where $K_r^a(h)$ denotes the cumulant generating function evaluated using the probability density $p^a(\bm{x})$,
\begin{align}
K_r^a(h) = \ln \int d\bm{x} \ e^{h r(\bm{x})} p^a(\bm{x}) .
\end{align}
This inequality relates the KL divergence between two arbitrary probability densities $p^b(\bm{x})$ and $p^a(\bm{x})$ to the average of some quantity with respect to $p^b(\bm{x})$ and the cumulant generating function with respect to $p^a(\bm{x})$.
For the the continuous time-reversal operation, we can choose $p^b = \mathbb{P}^{\theta_1}$ and $p^a = \mathbb{P}^{\theta_2}$, which results in the inequality
\begin{align}
D_\text{KL}\big(\mathbb{P}^{\theta_1}[\hat{\bm{x}}] \Vert \mathbb{P}^{\theta_2}[\hat{\bm{x}}]\big) &\geq \sup_{h} \Big( h\av{J_\tau}^{\theta_1} - K_{J_\tau}^{\theta_2}(h) \Big) \label{kullback-current} \\
\Leftrightarrow \qquad \frac{(\theta_1 - \theta_2)^2}{4} S^\text{irr}_\tau &\geq \sup_{h} \Big( h \theta_1 \av{J_\tau} - K_{J_\tau}^{\theta_2}(h) \Big) \n ,
\end{align}
where we used the explicit expression \eqref{forward-forward-1} for the KL divergence and the scaling of the average current with $\theta_1$.
Since the inequality holds for any $h$, we may rewrite this as a bound on the cumulant generating function
\begin{align}
K_{J_\tau}^{\theta_2}(h) \geq h \theta_1 \av{J_\tau} - \frac{1}{4} \big(\theta_1 - \theta_2 \big)^2 \Delta S^\text{irr}_\tau,
\end{align}
Since the left-hand side is independent of $\theta_1$, we may maximize the right-hand side with respect to $\theta_1$ and obtain
\begin{align}
K_{J_\tau}^{\theta_2}(h) \geq h \theta_2 \av{J_\tau} + \frac{h^2 \av{J_\tau}^2}{\Delta S^\text{irr}_\tau},
\end{align}
which is the desired quadratic lower bound.
Note that this can be written in a more compact way by introducing the cumulant generating function of the current fluctuations $\Delta J^\theta_\tau[\hat{\bm{x}}] = J_\tau[\hat{\bm{x}}] - \av{J_\tau}^\theta$,
\begin{align}
K_{\Delta J^\theta_\tau}^{\theta}(h) \geq \frac{h^2 \av{J_\tau}^2}{\Delta S_{\text{irr},\tau}} .
\end{align}
Since the right-hand side is independent of $\theta$, this implies that the fluctuations of $J_\tau$ for any $\theta$ are governed by a common lower bound.
In particular, expanding the cumulant generating function for small $h$, we recover \eqref{main-TUR-theta} of the main text,
\begin{align}
\text{Var}^\theta(J_\tau) \geq \frac{2 \av{J_\tau}^2}{\Delta S_{\text{irr},\tau}},
\end{align}
which represents a $\theta$-independent lower bound on the variance of $J_\tau$ in the dynamics at parameter value $\theta$.

We may also use \eqref{kullback-current} as a lower bound on the entropy production,
\begin{align}
\Delta S^\text{irr}_\tau &\geq 4 \sup_{h} \bigg( \frac{h \theta_1 \av{J_\tau} - K_{J_\tau}^{\theta_2}(h)}{(\theta_1- \theta_2)^2} \bigg).
\end{align}
Again maximizing with respect to $\theta_1$, we find
\begin{align}
\Delta S^\text{irr}_\tau \geq \sup_{h,\theta_2} \bigg( \frac{h^2 \big(\av{J_\tau} \big)^2 }{K_{J_\tau}^{\theta_2}(h) - h \theta_2 \av{J_\tau}} \bigg) = \sup_{h,\theta_2} \bigg( \frac{h^2 \big(\av{J_\tau} \big)^2 }{K_{\Delta J_\tau^{\theta_2}}^{\theta_2}(h)} \bigg) \label{entropy-bound-0}.
\end{align}
On the other hand, we may also obtain the equivalent of \eqref{kullback-current} for the discrete time-reversal operation by choosing $\theta_1 = 1$ and $\theta_2 = -1$, which yields see also Ref.~\cite{Dec20}
\begin{align}
\Delta S^\text{irr}_\tau \geq \sup_{h} \Big(h \av{J_\tau} - K_{J_\tau}(-h) \Big) = \sup_{h} \Big(2 h \av{J_\tau} - K_{\Delta J_\tau}(-h) \Big),
\end{align}
where we used that $\theta = -1$ is the time-reverse of $\theta = 1$ and thus $K_{J_\tau}^{-1}(h) = K_{J_\tau}^{1}(-h)$.
Setting $\theta_2 = -1$ in \eqref{entropy-bound-0}, we thus have the pair of bounds
\begin{align}
\Delta S^\text{irr}_\tau \geq \left\lbrace \begin{array}{l} \sup_{h} \bigg( \frac{ h^2 \big(\av{J_\tau} \big)^2 }{K_{\Delta J_\tau}(-h)} \bigg) \\[2ex] \sup_{h} \Big(2 h \av{J_\tau} - K_{\Delta J_\tau}(-h) \Big) . \end{array} \right.
\end{align}
We further have
\begin{align}
\frac{ h^2 \big(\av{J_\tau} \big)^2 }{K_{\Delta J_\tau}(-h)} - 2 h \av{J_\tau} + K_{\Delta J_\tau}(-h) = \frac{1}{K_{\Delta J_\tau}(-h)} \Big( h \av{J_\tau} - K_{\Delta J_\tau}(-h) \Big)^2 \geq 0 .
\end{align}
Thus, for any $h$, the bound from the continuous time-reversal operation is tighter, i.~e., we have
\begin{align}
\Delta S^\text{irr}_\tau \geq \sup_{h} \bigg( \frac{ h^2 \big(\av{J_\tau} \big)^2 }{K_{\Delta J_\tau}(-h)} \bigg) \geq \sup_{h} \Big(2 h \av{J_\tau} - K_{\Delta J_\tau}(-h) \Big).
\end{align}
This shows that, indeed, the continuous time-reversal operation allows us to obtain a tighter bound on the entropy production than the usual, discrete time-reversal.
Note that the above bound also constitutes an extended version of the TUR involving higher-order cumulants of the current,
\begin{align}
\Delta S^\text{irr}_\tau \geq \sup_{h} \bigg( \frac{ h^2 \big(\av{J_\tau} \big)^2 }{K_{\Delta J_\tau}(h)} \bigg), \label{TUR-nonlinear}
\end{align}
which reduces to the TUR in the limit $h \rightarrow 0$.
For the entropy production $J_\tau = \Sigma_\tau$, we have $K_{\Delta \Sigma_\tau}(-1) = K_{\Sigma_\tau}(-1) + \Delta S^\text{irr}_\tau$ and
\begin{align}
K_{\Sigma_\tau}(-1) = \ln \Av{e^{-\Sigma_\tau}} = \ln 1 = 0
\end{align}
from the fluctuation theorem.
Thus, \eqref{TUR-nonlinear} is an equality for $J_\tau = \Sigma_\tau$ and $h = -1$.
Comparing this to the RTUR, this suggest that either, we may either consider the fluctuations the current relative to its local mean, which are Gaussian for the stochastic entropy production, or we may take into account the higher order cumulants via \eqref{TUR-nonlinear}; in both cases, the respective inequality turns into an equality when choosing the stochastic entropy production as an observable.

\section{Diffusion in a tilted periodic potential}
As discussed in the main text, the motion of an overdamped particle in a tilted periodic potential is described by the Langevin equation
\begin{align}
\dot{x}(t) = \mu (- U'(x) + F) + \sqrt{2 \mu T} \xi(t) .
\end{align}
The corresponding steady-state Fokker-Planck equation reads
\begin{align}
0 = - \partial_x \big( \nu_\text{st}(x) p_\text{st}(x) \big) .
\end{align}
In the one-dimensional case, the only possible solution is $\nu_\text{st}(x) p_\text{st}(x) = \omega_0$, where $\omega_0$ is a constant with dimensions of frequency.
Using the explicit expression for the local mean velocity
\begin{align}
\nu_\text{st}(x) = \mu ( U'(x) + F ) - \mu T \partial_x \ln p_\text{st}(x),
\end{align}
this yields the first-order differential equation for $p_\text{st}(x)$,
\begin{align}
\mu ( -U'(x) + F ) p_\text{st}(\bm{x}) - \mu T \partial_x p_\text{st}(x) = \omega_0
\end{align}
with the general solution
\begin{align}
p_\text{st}(x) = e^{-\frac{U(x)-F x}{T}} \bigg( c - \frac{\omega_0}{\mu T} \int_{0}^x dy \ e^{\frac{U(y) - F y}{T}} \bigg) .
\end{align}
The constants $c$ and $\omega_0$ are determined by the conditions that $p_\text{st}(x)$ should be periodic, $p_\text{st}(x + L) = p_\text{st}(x)$, and normalized, $\int_0^L dx \ p_\text{st}(x) = 1$.
Solving these conditions for $c$ and $\omega_0$ yields \cite{Rei01}
\begin{align}
p_\text{st}(x) = \frac{e^{-\frac{U(x) - F x}{T}} \int_x^{x+L} dy \ e^{\frac{U(y) - F y}{T}}}{\int_0^L dx \ e^{-\frac{U(x)- F x}{T}} \int_x^{x+L} dy \ e^{\frac{U(y) - F y}{T}}} \label{perpot-pdf} 
\end{align}
and 
\begin{align}
\nu_\text{st}(x) = \mu T \big(1 - e^{- \frac{F L}{T}} \big) \frac{e^{\frac{U(x)-F x}{T}}}{\int_{x}^{x+L} dy \ e^{\frac{U(y)-F y}{T}}}. \label{perpot-meanvel}
\end{align}

\end{document}